%% file: main.tex
\newif\iflncs
\lncsfalse

\newif\ifshowcomments

\iflncs
\input{ECpreamble}
\else 
\input{preamble}
\fi

\title{Public-Key Encryption from the MinRank Problem}

\iflncs
\author{}
\institute{}
\date{}
\else
\author{Rohit Chatterjee\thanks{\orcidlink{0009-0004-1970-1779}
 Department of Computer Science, National University of Singapore. Email: \href{mailto:rochat@nus.edu.sg}{rochat@nus.edu.sg}} \and Changrui Mu\thanks{\orcidlink{0009-0008-4945-9353} Department of Computer Science, National University of Singapore; School of Computer Science, Carnegie Mellon University. Email: \href{mailto:changrui.mu@u.nus.edu}{changrui.mu@u.nus.edu}} \and Prashant Nalini Vasudevan\thanks{\orcidlink{0000-0001-6880-795X} Department of Computer Science, National University of Singapore. Email: \href{mailto:prashant@comp.nus.edu.sg}{prashvas@nus.edu.sg}}}
\date{\today}
\fi

\begin{document}

\pagenumbering{roman}

\maketitle              

\thispagestyle{empty}

\input{abstract}

\setcounter{tocdepth}{2}
\iflncs
\else
\tableofcontents
\fi

\newpage
\pagenumbering{arabic}

\input{intro}
\input{prelim}
\input{assm}

\input{pke}

\input{analysis.tex}

\iflncs
\else
\subsection*{Acknowledgements}
This work was supported by the National Research Foundation, Singapore, under its NRF Fellowship programme, award no. NRF-NRFF14-2022-0010. \pnote{Changrui, fill in funding info here} AI tools were used as typing assistants for grammar and basic editing, and for mild assistance with LaTeX formatting.
\fi

\bibliographystyle{alpha}
\bibliography{references}

\appendix

\input{history}


\end{document}

%% file: preamble.tex
\documentclass[11pt]{article}

\usepackage{amsthm}
\usepackage{fullpage}

\usepackage{amsfonts}
\usepackage{amsmath}
\usepackage{mdframed}
\usepackage{float}
\usepackage{enumitem}
\usepackage[colorlinks=true,linkcolor=blue,citecolor=blue]{hyperref}
\usepackage{graphicx}
\usepackage{tikz}
\usepackage{xcolor} 
\usepackage{orcidlink}
\usepackage{thm-restate}
\usepackage[most]{tcolorbox}

\usepackage[capitalize]{cleveref}  
\usepackage{caption}
\usepackage{threeparttable}

\usepackage[margin=1in]{geometry}
\usepackage[T1]{fontenc}
\usepackage[utf8]{inputenc}
\usepackage{lmodern,microtype}
\usepackage{tikz}
\usetikzlibrary{arrows.meta,positioning}
\tikzset{
  box/.style={draw,rounded corners,align=left,font=\scriptsize,
              text width=40mm,inner sep=2pt},
  elabel/.style={midway,sloped,above,fill=white,inner sep=1pt},
}

\newcommand{\remove}[1]{}

\ifshowcomments
\newcommand{\cnote}[1]{[{\footnotesize \textsf{\color{orange}{\bf Changrui:} { {#1}}}}]}
\newcommand{\pnote}[1]{[{\footnotesize \textsf{\color{cyan}{\bf Prashant:} { {#1}}}}]}
\newcommand{\rnote}[1]{[{\footnotesize \textsf{\color{magenta}{\bf Rohit:} { {#1}}}}]}
\else
\newcommand{\cnote}[1]{}
\newcommand{\pnote}[1]{}
\newcommand{\rnote}[1]{}
\fi

\newtheorem{theorem}{Theorem}[section]
\newtheorem{conjecture}[theorem]{Conjecture}
\newtheorem{claim}[theorem]{Claim}
\newtheorem{corollary}[theorem]{Corollary}
\newtheorem{fact}[theorem]{Fact}
\newtheorem{lemma}[theorem]{Lemma}

\newtheorem{definition}[theorem]{Definition}
\theoremstyle{remark}
\newtheorem{remark}[theorem]{Remark}
\theoremstyle{plain}

\newenvironment{proofof}[1]{\begin{proof}[\textit{Proof of #1}]}{\end{proof}}

\crefname{definition}{Definition}{Definitions}
\crefname{sub-definition}{Definition}{Definitions}
\crefname{example}{Example}{Examples}
\crefname{exercise}{Exercise}{Exercises}
\crefname{property}{Property}{Properties}
\crefname{question}{Question}{Questions}
\crefname{solution}{Solution}{Solutions}
\crefname{theorem}{Theorem}{Theorems}
\crefname{informaltheorem}{Theorem}{Theorems}
\crefname{proposition}{Proposition}{Propositions}
\crefname{problem}{Problem}{Problems}
\crefname{lemma}{Lemma}{Lemmas}
\crefname{conjecture}{Conjecture}{Conjectures}
\crefname{corollary}{Corollary}{Corollaries}
\crefname{fact}{Fact}{Facts}
\crefname{claim}{Claim}{Claims}
\crefname{remark}{Remark}{Remarks}
\crefname{note}{Note}{Notes}
\crefname{figure}{Figure}{Figure}
\crefname{case}{Case}{Cases}
\crefname{proofsketch}{Proof Sketch}{Proof Sketches}

\newcommand{\eps}{\varepsilon}

\newcommand\pgets{\mathrel{\stackrel{\makebox[0pt]{\mbox{\normalfont\tiny \$}}}{\gets}}}

\def\ddefloop#1{\ifx\ddefloop#1\else\ddef{#1}\expandafter\ddefloop\fi}

\def\ddef#1{\expandafter\def\csname bb#1\endcsname{\ensuremath{\mathbb{#1}}}}
\ddefloop ABCDEFGHIJKLMNOPQRSTUVWXYZ\ddefloop

\def\ddef#1{\expandafter\def\csname c#1\endcsname{\ensuremath{\mathcal{#1}}}}
\ddefloop ABCDEFGHIJKLMNOPQRSTUVWXYZ\ddefloop

\newcommand{\pr}[1]         {\Pr\left[ #1 \right]}
\newcommand{\prob}[2]       {\Pr_{#1}\left[ #2 \right]}

\definecolor{mygreen}{RGB}{0,128,0} 
\newcommand{\resolved}[1]{{\color{mygreen}{[Resolved]}}}

\definecolor{linkwarm}{HTML}{8A2D0C}
\hypersetup{
  colorlinks=true,
  linkcolor=linkwarm,
  citecolor=linkwarm,
  urlcolor=linkwarm
}

\newcommand{\keygen}{\mathsf{KeyGen}}
\newcommand{\enc}{\mathsf{Enc}}
\newcommand{\dec}{\mathsf{Dec}}

\newcommand{\ip}[1]{\langle #1 \rangle}

\newcommand{\poly}{\mathrm{poly}}

\newcommand{\abs}[1]        {\left| #1\right|}

\newcommand{\ceil}[1]       {\left\lceil #1 \right\rceil}

\newcommand{\algA}{\mathsf{A}}
\newcommand{\algB}{\mathsf{B}}
\newcommand{\algR}{\mathsf{R}}

\newcommand{\td}{s}

\newcommand{\negl}{\mathrm{negl}}

\newcommand{\Fq}{\mathbb{F}_\q}
\newcommand{\q}{2}

\newcommand{\ignore}[1]{}
\usepackage{framed}
\usepackage{xcolor}

\newcommand{\groebner}{Gr\"obner~}

\newtcolorbox
[auto counter,
Crefname={Constr.}{Constr.},
]{ConstructionBox}[2][]
{
  title    = {},
  fonttitle=\small\bfseries,
  enforce breakable, 
  colframe = black,
  coltitle = black,
  colbacktitle = white, 
  coltext  = black, 
  colback  = white,
  arc=0pt,
  outer arc=0pt,
  boxrule=0.5pt,
  left=0pt, 
  right=0pt, 
   top=0pt,
  bottom=0pt,
}

\newtcolorbox
[auto counter,
Crefname={Prot.}{Prot.}
]{ProtocolBox}[2][]
{
  title    = Protocol~\thetcbcounter: #2,#1,
  fonttitle=\small\bfseries,
  enforce breakable, 
  colframe = black,
  coltitle = black,
  colbacktitle = white, 
  coltext  = black, 
  colback  = white,
  arc=0pt,
  outer arc=0pt,
  boxrule=0.5pt,
  toprule at break = 0pt,
  bottomrule at break = 0pt,
  left=0pt, 
  right=0pt, 
   top=0pt,
  bottom=0pt,
}


%% file: abstract.tex
\iflncs
\else

\begin{abstract}
We construct a public-key encryption scheme from the hardness of the (planted) MinRank problem over uniformly random instances. This corresponds to the hardness of decoding random linear rank-metric codes. Existing constructions of public-key encryption from such problems require hardness for structured instances arising from the masking of efficiently decodable codes. Central to our construction is the development of a new notion of duality for rank-metric codes.
\end{abstract}

\fi

%% file: intro.tex
\section{Introduction}
\label{sec:intro}

The MinRank problem is one of considerable significance to cryptography. An instance of this problem consists of $k$ matrices over $\mathbb{F}_2$, each of dimension $n \times n$, and the task is to find a linear combination of these matrices that has rank less than a given number $r$, promised that it exists. This problem was first explicitly defined by Buss et al. \cite{BFS99}, who also showed it to be NP-hard. Around the same time, Kipnis and Shamir~\cite{KS99} studied this problem in their cryptanalysis of the HFE cryptosystem, and showed an algorithm that runs in polynomial time if $n \geq \Omega(k\cdot r)$.

Since then, a long line of work has discovered that this problem arises naturally in the cryptanalysis of a wide variety of cryptographic schemes: key examples include the previously mentioned HFE and Rainbow cryptosystems (analysed in \cite{KS99,Beullens22}), the TTM~\cite{moh1999public} and GeMMs~\cite{casanova2020gemss} cryptosystems (analysed in \cite{goubinCryptanalysisTTMCryptosystem2000,BaenaBCPSV22}), etc.. Consequently, the complexity of the MinRank problem has been extensively studied over the decades~\cite{KS99,goubinCryptanalysisTTMCryptosystem2000,faugereCryptanalysisMinRank2008,FEDS10,FDS13,bardetImprovementsAlgebraicAttacks2020a,BaenaBCPSV22,BardetB22,BG25}. Various categories of attacks exist utilizing algebraic and combinatorial methods, and these have been analysed quite rigorously, and their practical runtimes have also been comprehensively studied (\cite{bardetImprovementsAlgebraicAttacks2020a} is a representative example). Overall, while efficient algorithms exist for some special cases like $n = \Omega(k\cdot r)$ and $r = O(1)$, the current consensus is that MinRank is hard to solve for a wide range of parameters, despite successive cryptanalytic advances. Further, no quantum attacks with significant quantum speedup are known either, making the problem plausibly quantum-hard (see \cite{AdjBBERSVZ24} for more on this).\footnote{For a more extensive discussion of the history of MinRank in cryptography, see \Cref{sec:history}. For a discussion of the known algorithms for the problem, see \Cref{sec:attacks}.}

\paragraph{Cryptography from MinRank.} This comprehensive study and the resulting belief in its hardness, together with its simplicity and linear structure, makes MinRank a promising starting point for constructing cryptosystems. Indeed, even early on, Courtois~\cite{Cou01} presented an identification scheme based on the hardness of MinRank, as well as a zero knowledge protocol and signature scheme. Subsequently, a burgeoning line of work~\cite{SantosoINY22,MD22,AdjRV23,Feneuil24,AdjBBERSVZ24,ABCFGNR23,adj2024mirath} has constructed identification and signature (as well as ring signature) schemes with improved parameters and additional properties, with some of them having been entered into the NIST standardization competition for post-quantum signatures~\cite{nist2016pqcstandardization}. 

\paragraph{Public-Key Encryption from MinRank.} There have also been multiple proposals of Public-Key Encryption (PKE) schemes based on the hardness of the MinRank problem. In this context, the problem is more naturally interpreted as the decoding problem for linear codes in the rank metric~\cite{Delsarte78,Gabidulin85}. In this problem, the instance is a tuple $(A_1,\dots,A_k)$ of $k$ matrices over $\bbF_2$ of dimension $n\times n$, together with another matrix $Y = \sum_i x_i\cdot A_i + E$ for some $x_1,\dots,x_k\in\bbF_2$ and a matrix $E$ of rank at most $r$. Given these, the task is to recover the $x_i$'s. The code here is the $\bbF_2$-linear space generated by the $A_i$'s, and $Y$ is a noisy codeword, with $E$ being the noise. Clearly this is an instance of the MinRank problem. Further, for the natural uniform distributions over their instances, these problems can be seen to be equivalent.


Most existing proposals for PKE related to MinRank follow the McEliece paradigm~\cite{McE78}, which has been considerably successful with codes in the Hamming metric~\cite{McE78,Nie86,BIKE17,HQC22}. Starting with the work of Gabidulin et al.~\cite{GPT91}, a series of such PKE schemes have been proposed, most of them relying on codes related to Gabidulin codes \cite{McEliece-PKE-gabidulinModifiedGPTPKC2001,McEliece-PKE-gabidulinReducibleRankCodes2003,McEliece-PKE-ourivskiColumnScramblerGPT2003a,McEliece-PKE-gabidulinAttacksCounterattacksGPT2008,McEliece-PKE-gabidulinErrorErasureCorrecting2008,McEliece-PKE-gabidulinImprovingSecurityGPT2009,McEliece-PKE-loidreauDesigningRankMetric2010,McEliece-PKE-rashwanSecurityGPTCryptosystem2011,McEliece-PKE-biglieriImpactIndependenceAssumptions2016,McEliece-PKE-wachter-zehRepairingFaureLoidreauPublicKey2018,McEliece-PKE-aragonMinRankGabidulinEncryption2024,McEliece-PKE-ModifiedGPTPKC2025,McEliece-PKE-GabidulinMatrixCodes,McEliece-PKE-gabidulinPublickeyCryptosystemsBased,McEliece-PKE-loidreauEvolutionGPTCryptosystem,GMRZ13}. While it can result in very efficient cryptosystems, a significant challenge with the McEliece paradigm is that it relies on the hardness of decoding codes that possess considerable structure. This has been a repeated concern in the above line of work, with many of the proposals being broken, subsequently repaired, broken again, and so on~\cite{gibsonSeverelyDentingGabidulin1995,overbeckNewStructuralAttack2005,ExtendingGibsonsAttacksa,overbeckStructuralAttacksPublic2008a,McEliece-PKE-horlemann-trautmannConsiderationsRankbasedCryptosystems2016,horlemann-trautmannExtensionOverbecksAttack2016,otmaniImprovedCryptanalysisRank2017,coggiaSecurityLoidreausRank2020,couvreurExtensionOverbecksAttack2023,porwalImprovedKeyAttack2025}. There have been a few proposals that do not fall into this paradigm~\cite{FL06,RP20,BBBG22}, but these too end up relying on the hardness of decoding structured codes, and have been subject to resultant attacks~\cite{GRS13}. (The survey by Bartz et al.~\cite{bartzRankMetricCodesTheir2022} is an excellent reference on this topic.)

These repeated attacks, while not uncommon in cryptographic research, highlight the risks inherent in relying on the hardness of problems with too much structure. Our objective in this paper is to instead construct a PKE scheme whose security follows solely from the hardness of decoding random, and thus unstructured, rank-metric codes. Or equivalently, from the hardness of MinRank on generic instances, which has remained broadly intractable in spite of decades of cryptanalysis.

\subsection{Our Results}
\label{sec:results}

We construct a public-key encryption scheme whose security follows from the hardness of the MinRank problem on uniformly random instances, or equivalently, the hardness of decoding random linear rank-metric codes from random noise. To be more precise, consider sampling uniformly random matrices $A_1,\dots,A_k\gets\bbF_2^{n\times n}$, field elements $x_1,\dots,x_k\gets\bbF_2$, and a matrix $E\in\bbF_2^{n\times n}$ that is uniform conditioned on having rank at most $r$. The complexity of breaking our encryption scheme is tightly related to that of recovering $x$ given $(A_1,\dots,A_k)$ and $Y = \sum_i x_i\cdot A_i + E$.

Our construction is presented in \cref{sec:pke}, together with the statement and proof of security, and discussions of the setting of the parameters $k$ and $r$ in relation to $n$. We survey the best potential attacks against our scheme in \cref{sec:attacks}. In our security proof, we need the hardness of a decisional version of the MinRank problem, which we derive from the hardness of the above search version using a search-to-decision reduction that we present in \cref{sec:s-to-d}. Our primary technical innovation is the development of a dual problem for MinRank based on a new duality notion for rank-metric codes, and showing that it is equivalent to MinRank in complexity. This is presented in \cref{sec:dual}, and discussed briefly below.

\paragraph{Non-Scalar Inner Products.} Our construction of PKE follows the approach of Alekhnovich's construction of PKE from the hardness of decoding random linear codes in the Hamming metric~\cite{Ale11}. This approach requires one crucial component -- an inner product that indicates whether its inputs are small. Suppose we are working with an ambient discrete vector space $\cX$ with some norm defined on it, in which our codewords exist. Then we need an inner product $\ip{\cdot, \cdot}:\cX\times\cX\rightarrow\cY$, for some space $\cY$, such that $\ip{a,b}$ behaves one way if both $a$ and $b$ are random \emph{small} elements from $\cX$ (according to the norm being used), and behaves in a different, efficiently distinguishable, way if either of $a$ or $b$ is uniformly random over $\cX$.

In Alekhnovich's original setting, the space $\cX$ is the vector space $\bbF_2^n$ with the norm of a vector being its Hamming weight, and $\ip{\cdot,\cdot}$ is just the standard vector inner product over $\bbF_2$. If $a$ or $b$ is sampled uniformly at random from $\bbF_2^n$, then $\ip{a,b}$ is very close to being a random bit. Whereas if they were both sampled as random vectors of Hamming weight less than $\sqrt{n}$, then $\ip{a,b}$ has a noticeable bias towards $0$. This satisfies the conditions above.

In our case, the space $\cX$ is the vector space $\bbF_2^{n\times n}$, with the norm of a matrix being its rank. It is not clear, however, whether the kind of inner product we need exists here. One natural candidate to try is the Frobenius inner product: $\ip{A,B}_F = Trace(A^TB)$, which outputs an element in $\bbF_2$. Another candidate is a product commonly used in the context of Gabidulin codes: $\ip{A,B}_G$ that treats $A$ and $B$ as vectors from $\bbF_{2^n}^n$ (taking each row to be an element of $\bbF_{2^n}$), and outputs the vector inner product between them over $\bbF_{2^n}$. Both products are close to being uniformly random if $A$ or $B$ is. If $A$ and $B$ are taken to be random matrices of somewhat low rank, both of these inner products are indeed biased towards $0$, but the bias is not large enough to be efficiently detected. So these do not fit our conditions.\footnote{Nevertheless, there is a substantially well-developed theory of duality of rank-metric codes based on these inner products~\cite{Ravagnani16,Gorla21}.}

Instead, we define a new inner product parametrized by some $t\in\bbN$ such that $t$ divides $n$. This inner product $\ip{A,B}_t$ outputs a $t\times t$ matrix over $\bbF_2$, and is computed by first dividing up $A$ and $B$ into $t\times t$ blocks, each an $(n/t)\times (n/t)$ sub-matrix, in the natural manner. Then $(i,j)^{\text{th}}$ entry of $\ip{A,B}_t$ is set to be equal to the Frobenius product between the $(i,j)^{\text{th}}$ blocks of $A$ and $B$. We then prove that if both $A$ and $B$ have rank less than $\sqrt{t}$, then $\ip{A,B}_t$ has rank less than $t$. On the other hand, if $A$ or $B$ is a uniformly random matrix, then $\ip{A,B}_t$ has rank $t$ with high probability. This satisfies the properties needed. There are more general inner products that may be defined following this template, but this is the one we found to be simplest to use for the task described next.

\paragraph{Duality.} Once such an inner product is defined, one more ingredient is needed -- the computational hardness of a ``dual'' problem that is defined by it. In Alekhnovich's case, the ``primal'' problem was the decoding of random linear codes. Or rather, the decision version of this problem -- given random $a_1,\dots, a_k \gets \bbF_2^n$ and $y$ that is either uniformly random or $y= \sum_ix_i \cdot a_i + e$ for a random small vector $e$, decide which is the case. The dual problem here is, given random $a_1,\dots,a_k\gets\bbF_2^n$ and $z_1,\dots,z_k \in \bbF_2$ that are either all random, or set to $z_i = \ip{r,a_i}$ for a random small $r$, to decide which is the case. This is a decision version of the syndrome decoding problem, and to complete the construction it is required that this problem also be computationally hard. Alekhnovich was able to show that, for appropriate parameters, this dual problem is actually equivalent to the primal problem of decoding.

We do the same for the dual of the decoding problem for rank-metric codes with our block-wise inner product described above. Here, the dual problem is, given random $A_1,\dots,A_k\gets \bbF_2^{n\times n}$ and $Z_1,\dots,Z_k\in\bbF_2^{t\times t}$ that are either all random, or set to $Z_i = \ip{R,A_i}_t$ for a random low-rank $R$, to decide which is the case. This again is a syndrome decoding problem, with the syndrome defined using our inner product. We show that, for suitable parameters, the MinRank problem reduces to this problem.\footnote{In fact, our techniques can be used to show that this problem is equivalent in complexity to the MinRank problem, though we do not include the proof here.} Thus, just assuming that MinRank is hard (for the appropriate parameters) gives us all the hardness we need to complete the construction following Alekhnovich's approach.

\paragraph{Significance.} As mentioned above, we see the primary significance of our construction as its lack of reliance on the hardness of decoding structured codes. Additionally, we believe it is plausibly post-quantum-secure, due to the lack of non-trivial quantum algorithms for the MinRank problem so far. We would also like to highlight its simplicity -- the construction and proof only need elementary linear algebra, and should be easy to understand and implement. We believe our notion of matrix-valued inner products and the resulting duality of rank-metric codes would be interesting to study even outside the context our construction. We also hope that our brief survey of the complexity of various attacks on the MinRank problem will be useful as reference for future work in the area.

\subsection*{Organization}

\iflncs
\else 
\cnote{Modify if the structure is different}
The rest of the paper is organized as follows: we give necessary preliminaries and notation in \Cref{sec:prelims}. In \Cref{sec:minrank}, we define the MinRank problems we consider. We present our inner product and the duality of MinRank in \Cref{sec:dual}. We present our search-to-decision reduction for MinRank in \Cref{sec:s-to-d}. We describe our PKE scheme in \Cref{sec:pke}, along with the proof of security and a discussion of parameter settings. These depend on the cost of various known attacks on the MinRank problem, which we survey and describe in \Cref{sec:attacks}. 


\fi


%% file: prelim.tex
\section{Preliminaries}
\label{sec:prelims}

\paragraph{Notation.}
We write $x\pgets X$ to indicate that $x$ is sampled from a distribution $X$. When $X$ is a set, this means that $x$ is sampled uniformly from $X$. We focus on the binary field $\Fq$, and adopt $+$ and $-$ instead of $\oplus$ to get a more generalizable construction, which can be easily adapted to larger fields. Vectors are represented by lowercase letters (e.g., $a\in \Fq^n$), matrices by upper-case letters (e.g., $A\in \Fq^{m\times n}$), and sequences of matrices of the same size by bold upper-case letters (e.g., $\mathbf{A}=(A_1,\dots,A_k)$, with each $A_i\in \Fq^{m\times n}$). For a vector $v=(v_1,\dots, v_k)\in \Fq^k$ and a sequence of matrices $\mathbf{A}=(A_1,\dots,A_k) \in ({\Fq}^{m\times n})^k$, the linear combination of $\mathbf{A}$ specified by the vector $v$ is denoted by $\mathbf{A}(v)=\sum_{i\in [k]}v_i\cdot A_i$. 
By $E_{ij}\in \Fq^{n\times n}$, we denote the unit matrix with $1$ in the position $(i,j)$.  We reserve the symbol $I_{n}$ for representing the $n\times n$ identity matrix. 

Let ${\rm vec}(A)$ denote the vectorization of a matrix $A\in \Fq^{m\times n}$, i.e., the operation that stacks the columns of $A=(a_{ij})\in \Fq^{m\times n}$ on top of one another to form a single column vector $(a_{11},a_{21},\dots,a_{nm})^{\rm T}\in \Fq^{mn}$ (e.g., for $A\in \Fq^{2\times 2}$, ${\rm vec}(A)=(a_{11},a_{21},a_{12},a_{22})^{\rm T}$). 

For two distribution ensembles $X=\{X_n\}_{n\in \mathbb{N}}$ and $Y=\{Y_n\}_{n\in \mathbb{N}}$, We write $X \approx_c Y$ to represent that the two distributions are computationally indistinguishable (see, e.g., {\cite[Dfn 3.2.2]{goldreichFoundationsCryptography2004}}).


\subsection{Public key encryption}
\label{sec:prelims-pke}


    
    

We define the standard notion of public key encryption schemes. 

\begin{definition}[Public Key Encryption Scheme]
  A public-key encryption scheme (PKE) is a triple of polynomial-time algorithms $(\keygen, \enc, \dec)$:
  \begin{itemize}
    \item $\keygen (1^n)$: On input security parameter, output a public key $pk$ and secret key $sk$.
    \item $\enc( pk, x\in \{0,1\})$: On input a plaintext $x\in \{0,1\}$ and public key $pk$, encrypt $x$ using $pk$, and output a ciphertext $ct$.
    \item  $\dec( sk, ct)$: On input a ciphertext $ct$ and a secret key $sk$, decrypt $ct$ using $sk$, and output the decrypted message $x'$.
  \end{itemize}
  These algorithms are required to satisfy the following two properties:
  \begin{itemize}
    \item {\bf Correctness:} There is a negligible function $\negl$ such that for every $x\in\{0,1\}$ and $n\in\bbN$:
    \begin{align*}
      \pr{\dec( sk, \enc(pk,x)) = x\ \Big|\ (pk,sk)\gets\keygen(1^n)} \geq 1-\negl(n).
    \end{align*}
    
    \item {\bf Semantic Security:} Encryptions of $0$ and $1$ should be computationally indistinguishable. That is, for every polynomial-time algorithm $\algB$, there is a negligible function $\negl$ such that for all $n\in \bbN$:
    \begin{align*}
      \abs{\pr{\algB(pk,ct)=1\ \Big|\ \substack{(pk,sk)\leftarrow \keygen(1^n)\\ ct\leftarrow \enc(pk,0)}} - \pr{\algB(pk,ct)=1\ \Big|\ \substack{(pk,sk)\leftarrow \keygen(1^n)\\ ct\leftarrow \enc(pk,1)}}} \\ \leq \negl(n).
    \end{align*}
    We say that an algorithm $\algB$ breaks the security of the scheme if the above difference in probabilities is non-negligible.
  \end{itemize}
\end{definition}

\subsection{Coding Theory}
\label{sec:prelims-coding}

Our construction will rely on notions of linear error correcting codes defined over matrices, where the associated metric will be the {\em rank} of the difference of two matrices --- these are known as {\em rank-metric codes} in the literature. In our setting, we will focus only on certain aspects of such codes. We define and develop some relevant concepts and notation in the following.

\begin{definition}[Rank Distance]
  For $m,n\in\bbN$, consider the vector space $\Fq^{m\times n}$ of matrices over the finite field $\Fq$. The \emph{rank distance} between any pair of matrices $A$ and $B$ is defined to be $d_{rank}(A,B) = rank(A-B)$.
\end{definition}

\begin{definition}[Matrix Code \cite{Hua51}]
    For $m,n\in\bbN$, consider the vector space $\Fq^{m\times n}$ over finite field $\Fq$. A matrix code $\mathcal{C}$ is a subset of $\Fq^{m\times n}$. The matrix code is linear if, in addition, it is a $\Fq$-subspace of $\Fq^{m\times n}$.
  \end{definition}

Throughout this paper, we will be dealing exclusively with linear matrix codes, and will simply refer to them as matrix codes after this section. The dual of a matrix code is defined with respect to a generalized notion of inner product that we define below.

\begin{definition}[Matrix-Valued Inner Product]
    \label{def:matrix-ip}
    Consider a finite field $\Fq$ and parameters $n,m,s,t\in\bbN$, a map $\langle\cdot, \cdot \rangle: \Fq^{m\times n} \times \Fq^{m\times n} \rightarrow \Fq^{t\times s}$ is a matrix-valued inner product if the following properties hold:
    \begin{itemize}
        \item \textbf{Symmetry}: For any matrices $A, B\in \Fq^{m\times n}$, $\langle A, B\rangle = \langle B,A \rangle$.
        \item \textbf{Bi-linearity}:
        $$\langle c_1A+c_2B,C\rangle = c_1\cdot \langle A,C\rangle + c_2\cdot \langle B,C\rangle,$$
        where $c_1,c_2\in \Fq$ are scalars. This should similarly hold in the second slot as well.
        \item \textbf{Non-degeneracy}: For any matrix $A\in \Fq^{m\times n}$, $\langle A, B\rangle = 0^{s\times t}$ for all $B\in \Fq^{m\times n}$ if and only if $A=0^{m\times n}$
    \end{itemize}
  We say two matrices $A$ and $B$ are \emph{orthogonal} to each other under $\langle \cdot,\cdot\rangle$ if $\langle A,B\rangle =0^{t\times s}$.
\end{definition}

\begin{definition}[Dual Matrix Code]
    For a finite field $\Fq$ and $m,n\in \mathbb{Z}_+$, consider a matrix code $\mathcal{C}\subseteq \Fq^{m\times n}$ over the finite field $\Fq$, and a matrix-valued inner product $\langle \cdot, \cdot \rangle: \Fq^{m\times n} \times \Fq^{m \times n}\rightarrow \Fq^{t\times s}$. The dual code of $\mathcal{C}$ with respect to the inner product, denoted by $\mathcal{C}^{\perp}$, is the subset of $\Fq^{m\times n}$ consisting of all matrices orthogonal to every matrix in $\mathcal{C}$ under $\langle \cdot, \cdot \rangle$:
    $$\mathcal{C}^{\perp}=\{H\in \Fq^{m \times n}~|~\langle H,C\rangle = 0^{s\times t}\text{ for all }C\in \mathcal{C}\}.$$
\end{definition}
It can be verified that for any inner product as above, the dual of a linear matrix code is also a linear matrix code. The most common inner product of matrices is the Frobenius product, where the output is a scalar, which can be viewed as a special case of the above with $t=s=1$.

\begin{definition}[Square Matrix Trace]
    Consider a finite field $\Fq$. The trace of a square matrix $A=(a_{ij})_{i,j\in [n]}\in \Fq^{n\times n}$ is the sum of their diagonal entries:
    $${\rm Tr}(A)=\sum\limits_{i\in [n]}a_{ii}.$$
\end{definition}

\begin{definition}[Frobenius Inner Product]
    Consider a finite field $\Fq$. For any two matrices $A=(a_{i,j}),B = (b_{i,j})\in \Fq^{m\times n}$, their Frobenius product is defined as:
    $$\langle A, B\rangle_F= \langle{\rm vec}(A), {\rm vec}(B)\rangle_F=\sum\limits_{i\in [m],j\in [n]} a_{ij}\cdot b_{ij} ={\rm Tr}(A^{\rm T}B),$$
\end{definition}


\begin{fact}[Rank vector/matrix independence~\cite{Coo05}]\label{fct:random matrix rank}
For integers $n\geq d \geq k \geq 1$ and a finite field $\Fq$, let $W\subseteq \Fq^n$ be any subspace of dimension $d$. Sample $k$ vectors $v_1,\dots, v_k \pgets W$ independently and uniformly from $W$, and form a $k\times n$ matrix $V=(v_1~|~\dots~|~v_k)^{\rm T}$. The probability that the vectors are independent is
$$\Pr[{\sf rank}(V)=k]= \prod\limits_{i=0}^{k-1} (1-\q^{i-d}).$$  
Consequently, $$\Pr[{\sf rank}(V)<k]\leq O(\q^{k-d}).$$
Furthermore, for each $s>0$, the probability that 
$$\Pr[{\sf rank}(V)=k-s]\leq O(2^{-s(s+d-k-1)}).$$
Consequently,
$$\Pr[{\sf rank}(V)<k-s]< (k-s)\cdot O(2^{-s(s+d-k-1)}),$$
where the $O(\cdot)$ hides constant dependent only on the finite field size,  independent of $n,d,k,s$.
\end{fact}

\begin{definition}[Total Variation Distance]
   For any two distributions $X$ and $Y$ supported on some set $\mathcal{S}$, their total variation distance is defined as:
   $$\Delta(X,Y)=\frac{1}{2}\cdot \sum\limits_{s\in \mathcal{S}} \left|\Pr[X=s]-\Pr[Y=s]\right|={\rm sup}_{S\subseteq \mathcal{S}}\left|\underset{s\pgets X}{\Pr}[s\in S]-\underset{s\pgets Y}{\Pr}[s\in S]\right|.$$
\end{definition}

%% file: assm.tex
\section{MinRank Problems}
\label{sec:minrank}

In this section we will look more closely at the MinRank problem and the associated hardness assumption. Our encryption scheme relies in particular on the  decisional variant of the MinRank problem on uniformly random instances.
We provide formulation of this problem and conjecture its hardness in some parameter regime, supported by the state-of-art attacks known to date. 

Our scheme is also based on the hardness of a ``dual'' variant of the MinRank problem, namely dual MinRank, defined by our new block-wise inner product. We formalize the block-wise inner product notion and show its intriguing rank-separating properties. 
We define the dual MinRank problem with respect to our new block-wise inner product and show that under a wide range of parameters, its hardness is comparable to the MinRank problem. 

Additionally, we also show a average-case search-to-decision reduction for generic uniform MinRank. To the best of our knowledge, it has not been formalized in the literature before. 




\subsection{The MinRank Problems}

In the search version of the MinRank problem, one is given $k+1$ matrices  $B_1,B_2,\dots, B_{k}, B_{k+1}$ of dimension $n\times n$, and asked to find a non-trivial linear combination of them such that the resulting matrix has rank less or equal to a prescribed parameter $r$. The MinRank problem was first introduced by Buss, Frandsen and Shallit~\cite{BFS99}. In the same paper, they showed that the MinRank problem over finite fields (e.g. $\Fq$) is  NP-hard.
In this work, we focus on matrices over a finite field $\Fq$, and the security of our construction relies on  average-case complexity of the decisional version of the problem. 


\begin{definition}[Decision MinRank Problem]\label{dfn:minrank dist}
Given polynomially bounded  $k=k(n)$, $r=r(n)$, consider a distribution of $k+1$ matrices sampled as follows for $n\in\mathbb{Z}_{+}$: 

$$\mathcal{D}_{{\rm MinRk}({n,k,r})}=\left(({\bf A},{\bf A}(s)+E)\;\Big|\; 
\begin{array}{l}
s\pgets \Fq^k;\; {\bf A}\pgets (\Fq^{n\times n})^k \\
E\pgets \Fq^{n\times n}\text{ s.t. }{\sf rank}(E)\leq r
\end{array}
\right).$$

The goal of the problem ${{\rm MinRk}_{n,k,r}}$ is to distinguish between the distribution $\mathcal{D}_{{\rm MinRk}({n,k,r})}$ and the uniform distribution $({\bf A}_{\rm R},U) \pgets (\Fq^{n\times n})^k\times \Fq^{n\times n}$. The \emph{advantage} of a distinguisher $\algA$ is defined as the absolute difference of its acceptance probability over the two distributions:

\iflncs
\else
The goal of the problem ${{\rm MinRk}_{n,k,r}}$ is to distinguish between the distribution $\mathcal{D}_{{\rm MinRk}({n,k,r})}$ and the uniform distribution $({\bf A}_{\rm R},U) \pgets (\Fq^{n\times n})^k\times \Fq^{n\times n}$. The \emph{advantage} of a distinguisher $\algA$ is defined as the absolute difference of its acceptance probability over the two distributions:
$${\rm Adv}_{{\rm MinRk}({n,k,r})}(\algA)=\left|\Pr\left[b=1 ~\Big|~ \substack{({\bf A}, {\bf A}(s)+E)\pgets \mathcal{D}_{{\rm MinRk}({n,k,r})}\\ b\leftarrow \algA\left({\bf A}, {\bf A}(s)+E\right)}\right]-\Pr\left[b=1  ~\Big|~ \substack{(\mathbf{A}_{\rm R},U) \pgets (\Fq^{n\times n})^k\times \Fq^{n\times n}\\ b\leftarrow \algA({\bf A}_{\rm R},U)}\right]\right|,$$
\fi
\end{definition}

In the Search MinRank problem, the task is instead to recover the linear combination $s$.

\begin{definition}[Search MinRank Problem]\label{dfn:minrank-planted-search}
    Given polynomially bounded  $k=k(n)$, $r=r(n)$, consider the distribution $\mathcal{D}_{{\rm MinRk}({n,k,r})}$ defined in \cref{dfn:minrank dist}. Given a sample $({\bf A},Y)$ from $\mathcal{D}_{{\rm MinRk}({n,k,r})}$, the goal of ${\rm SearchMinRk}_{n,k,r}$ is to find any $s'\in\Fq^k$ such that ${\sf rank}\!\left(Y-{\bf A}(s')\right)\le r$. 
The \emph{success probability} of an algorithm for this problem is the probability that it correctly finds such an $s'$.
\end{definition}



The most naive way to solve either MinRank problem is exhaustive search -- either over the solution vector $s\in \Fq^k$ directly or over all low-rank matrice $E$ with ${\rm rank}(E)\leq r$. Both strategies require superpolynomial time if $k=\omega(\log n)$ and $r=\omega(\log n)$.  Despite decades of work (e.g. \cite{KS99,FEDS10,bardetImprovementsAlgebraicAttacks2020a}) since its first proposal \cite{BFS99} and application in cryptography \cite{KS99}, all existing algorithms (classical and quantum) for these problems still run in super-polynomial time in the regime $r=\omega(\log n)$, $k=\omega(\log n)$ and $\frac{k\cdot r}{n+k}=\omega(\log n)$ (See \Cref{sec:security analysis} for details). The following is thus a reasonable conjecture regarding its complexity.
\begin{conjecture}\label{conjecture:decisional MinRank}
There exist polynomially bounded $k=k(n)$ and $r=r(n)$ such that ${\rm MinRk}_{n,k,r}$  is hard. That is, for any non-uniform probabilistic polynomial-time distinguishing algorithm $\algA$, its advantage on the decision MinRank problem ${\rm Adv}_{{\rm MinRk}({n,k,r})}(\algA)$ is a negligible function (in $n$). 
\end{conjecture}

We make some observations below addressing the choices we have made in our formulation. 

\begin{remark}
    We omit the exact procedure for sampling the uniformly random noise matrix $E$ of low rank $\leq r$ in our description of schemes. A simple method is to randomly sample a rank value $\rho\leq r$ according to the right probability mass, and then sample a random rank-$\rho$ matrix, which can be done in a standard manner. In practice, it's sufficient to simply sample a random matrix of rank $r$ for large enough $r$ and $n$. Following \Cref{fct:random matrix rank}, a random matrix of rank at most $r$ has rank $r$ with overwheming probability in $n$, this distribution is thus statistically close to the prescribed one.
\end{remark}

\cnote{Does it make sense to put the above remark to \Cref{sec:pke}?, because we do not sample $E$ in this section }\rnote{I think this is still a natural point to address in the context of the definition. Maybe you can add a reference to this in section 4 where you feel its most useful, for emphasis?}

\begin{remark}
    The prevalent formulation of the decisional MinRank problem in existing literature (e.g., \cite{FEDS10}) involves distinguishing a tuple of $k+1$ independently uniformly random matrices ${\bf U}=(U_1,\dots, U_{k+1})$ from a tuple ${\bf B}=(B_1,\dots, B_{k+1})$ sampled subject to the condition that their span contains a matrix $B^*$ of rank at most $r$. This formulation can be easily derived from the distribution given in \cref{dfn:minrank dist} by randomly permuting the matrices in the tuple.
    Looking ahead, the MinRank description  in \cref{dfn:minrank dist}  lends itself naturally to the definition of a {\em dual} version of the problem, which will play a pivotal role in our PKE construction. Moreover, this formulation of the MinRank problem is closer to the analogue of the hard decoding problem for matrix codes with the rank metric, and thus aligns more closely with our intuition of existing constructions based on hard decoding problems.
\end{remark}

\begin{remark}
Hereon we focus on the MinRank problem over square matrices for notational simplicity, but our results can be naturally generalized to the case where the matrices are not square. 
\end{remark}

\subsection{Dual MinRank Problem}
\label{sec:dual}
 Here we define our new block-wise inner product, and the dual  decisional MinRank problem using the block-wise inner product.

 We show the useful properties of the block-wise inner product, and provide an average-case reduction from dual MinRank to MinRank; we refer this as the \emph{duality} propety. 

\begin{definition}[$t$-block-wise inner product $\langle \cdot, \cdot \rangle_{t}$] Consider $t,n\in\mathbb{Z}_{+}$ such that $t$ divides $n$.
Let $A$ and $B$ be two $n\times n$ matrices over $\Fq$. Divide $A$ and $B$ into $t^2$ square matrices, each of dimension $(n/t)\times (n/t)$, as follows: 
$$A=\begin{pmatrix}
    A^{11}& A^{12}& \dots& A^{1t}\\
    A^{21}& A^{22}& \dots& A^{2t}\\
    \vdots & \vdots& \ddots& \vdots\\
    A^{t1}& A^{t2}& \dots& A^{tt}
\end{pmatrix}  \quad B=\begin{pmatrix}
    B^{11}& B^{12}& \dots& B^{1t}\\
    B^{21}& B^{22}& \dots& B^{2t}\\
    \vdots& \vdots& \ddots& \vdots\\
    B^{t1}& B^{t2}& \dots& B^{tt}
\end{pmatrix},$$
where each submatrix $A^{ij}, B^{ij} \in \Fq^{(n/t)\times (n/t)}$. The $t$-block-wise inner product of $A$ and $B$ is defined to be a matrix of dimension $t\times t$ as follows:
$$\langle A, B \rangle_t=\sum\limits_{i,j\in [t]} E_{ij}\; \langle A^{ij},B^{ij}\rangle_F=\begin{pmatrix}
    \langle A^{11}, B^{11}\rangle_F& \langle A^{12}, B^{12}\rangle_F& \dots& \langle A^{1t}, B^{1t}\rangle_F\\
    \langle A^{21}, B^{21}\rangle_F& \langle A^{22}, B^{22}\rangle_F& \dots& \langle A^{2t}, B^{2t}\rangle_F\\
    \vdots & \vdots& \ddots& \vdots\\
    \langle A^{t1}, B^{t1}\rangle_F& \langle A^{t2}, B^{t2}\rangle_F& \dots& \langle A^{tt}, B^{tt}\rangle_F\\
\end{pmatrix}, $$
where $\langle\cdot,\cdot\rangle_F$ is the Frobenius inner product.
\end{definition}

Following the corresponding properties of the Frobenius inner product, we show that our notion of the $t$-block-wise inner product is also a matrix-valued inner product over the field $\Fq$.

\begin{claim}
    For any $n,t\in \mathbb{Z}_+$ such that $t$ divides $n$, the $t$-block-wise inner product $\ip{\cdot,\cdot}_t:\Fq^{n\times n}\times\Fq^{n\times n}\rightarrow \Fq^{t\times t}$ satisfies the properties of Symmetry, Bilinearity, and Non-degeneracy required of matrix-valued inner products (in \cref{def:matrix-ip}).
\end{claim}

To simplify notation, given a sequence of matrices ${\bf A}=(A_1,A_2,\dots,A_k)$ and $B$ where all the $A_i$'s and $B$ are in $\Fq^{n\times n}$, we define the following shorthand notation: $$\langle {\bf A}, B \rangle_t=\left(\langle A_1, B \rangle_t,\langle A_2, B \rangle_t,\dots, \langle A_k, B \rangle_t\right)\in (\Fq^{t\times t})^{k}.$$

\iflncs
\else
\begin{definition}[Dual MinRank Problem]
    Consider polynomially bounded $l=l(n)$, $r=r(n)$, $t=t(n)$ such that $t|n$, the dual MinRank distribution is defined as:
    $$\mathcal{D}_{{\rm dualMinRk}({n,l,r})}=\left(({\bf H}, \langle{\bf H}, E\rangle_t )\;\Big|\; E\pgets \Fq^{n\times n}\text{ s.t. }{\sf rank}(E)\leq r,\; {\bf H}\pgets (\Fq^{n\times n})^{l}\right).$$
    The advantage of a distinguisher $\algA$ in solving the ${\rm dualMinRk}({n,l,r})$ problem is defined as the absolute difference of its acceptance probability on the dual MinRank distribution and the uniform distribution: 
    $${\rm Adv}_{{\rm dualMinRk}({n,l,r})}(\algA)=\left|\Pr\left[b=1~\Big|~ \substack{({\bf H}, \langle {\bf H}, E \rangle_t)\pgets \mathcal{D}_{{\rm dualMinRk}({n,l,r})}\\ b\leftarrow \algA\left({\bf H}, \langle {\bf H}, E \rangle_t\right)}\right]-\Pr\left[b=1~\Big|~ \substack{(\mathbf{H}_{\rm R},{\bf C}) \pgets (\Fq^{n\times n})^l\times (\Fq^{t\times t})^l\\ b\leftarrow \algA(\mathbf{H}_{\rm R},{\bf C})}\right]\right|,$$
\end{definition}
\fi

Next, we show that the dual MinRank problem is essentially as hard to solve as the decision MinRank problem. The security of our encryption scheme will rely critically on this property. 

\begin{lemma}[Duality]\label{lmm:reduce MinRank to dualMinRank}
Consider any polynomially bounded functions $r=r(n)$, $k=k(n)$, $l=l(n)$, and $t=t(n)$ that have the following properties:
\begin{itemize}[topsep=0pt]
    \item $(n/t)^2-k-l=\omega(\log n)$. 
    \item $t$ divides $n$.
\end{itemize}
 If there exists a distinguisher for ${\rm dualMinRk}({n,l,r})$ that runs in time $T(n)$ and has advantage $\epsilon(n)$, then there exists a distinguisher for ${\rm MinRk}({n,k,r})$ that runs in time $T(n)+\poly(n)$
 and has advantage $(\epsilon(n)-\negl(n))$.
\end{lemma}

The proof of this result will rely mainly on some useful properties of the blockwise inner product we defined previously. Thus before showing this proof, we first describe and show these properties. 

\subsubsection{Properties of the Block-Wise Inner Product}
\label{sec:block-ip}

The $t$-block-wise inner product can be equivalently defined using the Kronecker product. 

\begin{definition}[Kronecker product]
The Kronecker product of two matrices $A\in \Fq^{n_1\times m_1}$ and $B\in \Fq^{n_2\times m_2}$ is denoted as $A \otimes B$. The resulting matrix is of size $n_1n_2\times m_1m_2$:
\begin{align*}
    A\otimes B &= \left[\begin{matrix}
        a_{11}B & a_{12}B & \cdots & a_{1m_1}B\\
        a_{21}B & a_{22}B & \cdots & a_{2m_1}B\\
        \vdots & \vdots & \ddots & \vdots\\
        a_{n_1}B & a_{n_12}B & \cdots & a_{n_1m_1}B
    \end{matrix} \right].\\
\end{align*}
\end{definition}

\begin{fact}\label{fct:kronecker 2 vec}\label{fct:kronecker rank}\label{fct:vec 2 trace}
Consider positive integers $n,m,t,s\in \mathbb{Z}_+$, and suppose we have matrices $A,C\in \Fq^{m\times n}, X\in \Fq^{n\times t}, B\in \Fq^{t\times s}$. Then we have the following identities:
$${\rm vec}(AXB)= ({B^{\rm T} \otimes A}) \cdot {\rm vec}(X).$$
$${\rm rank}(A\otimes B)= {\rm rank}(A)\cdot  {\rm rank}(B).$$
$${\rm Tr}(A^{\rm T}C)={\rm vec}(A)^{\rm T}{\rm vec}(C).$$
\end{fact}



\begin{fact}[Shift Rule for Trace]\label{fct:shift}
Let $k\ge2$ and $A_1,\dots,A_k$ be conformable matrices so that $A_1\cdots A_k$ is square. Then the trace of their product is invariant under cyclic shift:
  $${\rm Tr}(A_1\cdots A_k)={\rm Tr}(A_2\cdots A_kA_1)=\cdots={\rm Tr}(A_kA_1\cdots A_{k-1}).$$
\end{fact}

\begin{claim}\label{clm:t-inner-prod-dfn2}
    For $r,n,t\in \mathbb{Z}_+$ such that  $t|n$ and $r:=n/t$, consider $A,B\in \Fq^{n\times n}$, let $E_{i,i} \in \Fq^{t\times t}$ be the matrix with a $1$ in $(i,i)$ and $0$ in other entries, and $I_{n/t}\in \Fq^{(n/t)\times (n/t)}$ be the identity matrix. Let $P_i = E_{ii}\otimes I_{n/t}$ and $p_i = {\rm vec}(P_i)$, and $P=(p_1~|~p_2~|~\dots~|~p_t)^{\rm T}$, then 
    $$\langle A, B\rangle_t = P\cdot (A \otimes B) \cdot P^{\rm T}.$$
\end{claim}
\begin{proof}
    For each $i,j\in[t]$, the $(i,j)$-th entry of $\langle A, B\rangle_t$ is 
    \begin{align}
       \langle A, B\rangle_t[i,j]= p_i^{\rm T} \cdot (A \otimes B)\cdot p_j &= {\rm vec}(P_i)^{\rm T} \cdot (A \otimes B)\cdot {\rm vec}(P_j) \\
       &={\rm vec}(P_i)^{\rm T}\cdot {\rm vec}\left(B  P_j  A^{\rm T}\right) \label{ln: kronecker 2 vec}\\
    &={\rm Tr}\left(P^{\rm T}_i B  P_j  A^{\rm T}\right)\label{ln:vec 2 trace}\\
    &={\rm Tr}\left(  P_j  A^{\rm T}P^{\rm T}_i B\right) \label{ln:shift} \\
   &={\rm vec}\left(  P_i  A P^{\rm T}_j\right)^{\rm T} \cdot {\rm vec}(B) \label{ln:trace 2 vec}\\
   &={\rm vec}\left(  A^{ij}\right)^{\rm T} \cdot {\rm vec}(B^{ij}) \label{ln:block retrieval}\\
   &= \langle A^{ij}, B^{ij} \rangle_F \label{ln:F product},
    \end{align}
    where line~(\ref{ln: kronecker 2 vec}) follows from \Cref{fct:kronecker 2 vec}; line~(\ref{ln:vec 2 trace}) and line~(\ref{ln:trace 2 vec}) follow from \cref{fct:vec 2 trace}; line~(\ref{ln:shift}) follows from \cref{fct:shift}; line~(\ref{ln:block retrieval}) follows from the definition of $P_i, P_j$; and line~(\ref{ln:F product}) follows from the definition of Frobenius product.
\end{proof}

Notice that ${\sf rank}(P)=t$ for $P$ defined in the claim above. From this equivalent definition, we can derive the following intriguing ``rank-preserving'' property of $t$-blockwise inner-product:

\begin{claim}
    For any matrices $A,B\in \Fq^{n\times n}$,
    $$ {\rm rank}\left(\langle A, B \rangle_t\right)\leq {\rm min}\left({\rm rank}(A)\cdot {\sf rank}(B), t\right).$$
\end{claim}
\begin{proof}
    \begin{align*}
        {\rm rank} \left( \langle A, B \rangle_t\right) \;&=\;  {\rm rank} \left(  P\cdot (A \otimes B)\cdot P^{\rm T}\right) \\
        &\leq \min\left( {\rm rank}(A\otimes B),\; {\rm rank}(P)\right)\\
        &=\min\left( {\rm rank}(A)\cdot  {\rm rank}(B),\; {\rm rank}(P)\right)\\
        &= {\rm min}\left({\sf rank}(A)\cdot {\sf rank}(B), t\right),
    \end{align*}
    where the second line follows from the fact that the column span of $ P\cdot (A \otimes B)$ must be a subspace of $P$'s column span, and the row span of $P\cdot (A \otimes B)$ must be a subspace of $(A \otimes B)$'s row span (i.e. Sylvester's inequality); and the third line follows from \Cref{fct:kronecker rank}.
\end{proof}

\subsubsection{Proof of Duality}
\label{sec:dual-proof}

We can now turn to proving \Cref{lmm:reduce MinRank to dualMinRank}. 

\begin{proofof}{\cref{lmm:reduce MinRank to dualMinRank}}
    Fix any values of $n$, $r$, $k$, $l$, and $t$ that satisfy the hypothesis of the theorem. Our approach is to construct a reduction algorithm that runs in time $O(k^{\omega-1}\cdot n^2 + l\cdot kn^2)$, and maps the distribution $\mathcal{D}_{{\rm MinRk}({n,k,r})}$ to $\mathcal{D}_{{\rm dualMinRk}({n,l,r})}$, and uniformly random $({\bf A_{\rm R}}, U)$ to uniformly random $({\bf H}_{\rm R}, V)$. This reduction would then immediately prove the theorem.
    
    For a sequence of matrices ${\bf A}=(A_1,\dots,A_k)$ where $A_i\in\Fq^{n\times n}$, define the $\langle\cdot,\cdot \rangle_t$-dual space of ${\bf A}$ as $${\bf A}^{\perp_t}=\{H\in \Fq^{n\times n}\;\big|\;\langle {\bf A}, H\rangle_t=(0^{t\times t})^k\}.$$ 
    Let ${\bf H}_{{\bf A}}$ and ${\bf H}_{\rm R}$ be the the following distributions:
        \begin{itemize}
            \item [] ${\bf H}_{\bf A}:$
            \item [] \begin{itemize}
                \item Samples a uniformly random $k$-tuple of matrices ${\bf A}=(A_1,\dots,A_{k})\pgets (\Fq^{n\times n})^{k}$.
                \item Samples $l$ random matrices uniformly from the $\langle\cdot,\cdot \rangle_t$-dual of ${\bf A}$: $$H_1,\dots, H_{l}\pgets {\bf A}^{\perp_t}.$$
                \item Outputs ${\bf H}=(H_1,\dots, H_{l})$.
            \end{itemize}
        \end{itemize}
        \begin{itemize}
            \item [] ${\bf H}_{\rm R}:$
            \item [] \begin{itemize}
                \item Samples $l$ random matrices uniformly $H_1,\dots, H_{l}\pgets \Fq^{n\times n}.$
                \item Outputs ${\bf H}=(H_1,\dots, H_{l})$.
            \end{itemize}
        \end{itemize}
    \begin{lemma}\label{clm:subspace to dual space}
        $\Delta\left( {\bf H}_{\bf A}, {\bf H}_{\rm R}\right)\leq t^2\cdot O(2^{k+l-(n/t)^2})\leq \negl(n)$.
    \end{lemma}
         We postpone the proof of \Cref{clm:subspace to dual space} to after the current proof of the theorem. Given an instance of decisional MinRank $({\bf A},Y)$, the reduction algorithm, which we call $\algR$, samples random $H_1,\dots, H_l \pgets {\bf A}^{\perp_t}$, which it can do in time $\poly(n)$.
         Then it sets ${\bf H}_{\bf A}=(H_1,\dots, H_l)$, and outputs $({\bf H}_{\bf A}, \langle {\bf H}_{\bf A}, Y\rangle_t)$.

         \medskip\medskip
         \noindent\underline{If $Y={\bf A}(s)+E$}, following the fact that ${\bf H}_{\bf A}\in {\bf A}^{\perp_t}$, we have $\langle{\bf H}_{\bf A}, Y\rangle_t = \langle{\bf H}_{\bf A}, {\bf A}(s)+E\rangle_t=\langle{\bf H}_{\bf A}, E\rangle_t$. By \cref{clm:subspace to dual space} and the data processing inequality,
            $$\Delta\left( ({\bf H}_{\bf A},\langle {\bf H}_{\bf A}, E\rangle_t),({\bf H}_{\rm R},\langle {\bf H}_{\rm R}, E\rangle_t) \right)\leq \negl(n),$$
            where ${\bf H}_{\bf A}\pgets ({\bf A}^{\perp_t})^l$, ${\bf H}_{\rm R}\pgets (\Fq^{n \times n})^l$, and $E\pgets \Fq^{n \times n}$ s.t. ${\rm rank}(E) \leq r$.

        \medskip\medskip
        \noindent\underline{If $Y\pgets \Fq^{n\times n}$}, we show that $({\bf H_A},\langle {\bf H_A},Y\rangle_t)$ is statistically close to the uniform distribution $({\bf H}_{\rm R}, V)\pgets (\Fq^{n\times n})^l \times (\Fq^{t\times t})^l$. By \cref{clm:subspace to dual space} and the data processing inequality,
             $$\Delta\left( ({\bf H}_{\bf A},\langle {\bf H}_{\bf A}, Y\rangle_t),({\bf H}_{\rm R},\langle {\bf H}_{\rm R}, Y\rangle_t) \right)\leq \negl(n),$$
             where $Y\pgets \Fq^{n\times n}$ is sampled randomly. Divide $Y$ and each $H_{z}\in {\bf H}_{\rm R}$ into $t^2$ sub-matrices of dimension $(n/t)\times (n/t)$ as follows:
        $$Y=\begin{pmatrix}
            Y^{11}& Y^{12}& \dots& Y^{1t}\\
            Y^{21}& Y^{22}& \dots& Y^{2t}\\
            \vdots & \vdots& \ddots& \vdots\\
            Y^{t1}& Y^{t2}& \dots& Y^{tt}
        \end{pmatrix}  \quad H_z=\begin{pmatrix}
            H_z^{11}& H_z^{12}& \dots& H_z^{1t}\\
            H_z^{21}& H_z^{22}& \dots& H_z^{2t}\\
            \vdots& \vdots& \ddots& \vdots\\
            H_z^{t1}& H_z^{t2}& \dots& H_z^{tt}
        \end{pmatrix},$$
        Because each $(i,j)$-th location in Y and ${\bf H}_{\rm R}$ is sampled independently, it's sufficient to fix any index $i,j\in [t]$, and then argue that $({\bf H}_{\rm R}^{ij}, \langle {\bf H}_{\rm R}^{ij}, Y^{ij} \rangle_F)$ is close to uniform, where $\langle {\bf H}_{\rm R}^{ij}, Y^{ij} \rangle_F=(\langle H_1^{ij}, Y^{ij} \rangle_F,\dots, \langle H_l^{ij}, Y^{ij} \rangle_F) \in \Fq^{\ell}$ is the Frobenius product. 
        
       \begin{lemma}[Leftover Hash Lemma]\label{lmm:lhl}
            Consider polynomially bounded  $n,m,k\in\mathbb{Z}_{+}$ and $\epsilon\in (0,1)$. Let $\mathcal{F}=\{f:\Fq^n \rightarrow \Fq^m\}$ be a family of pairwise independent hash functions and let $k= m+2\log (1/\epsilon)$, then for any distribution $X$ over $\Fq^n$ with min-entropy $H_{\infty}(X)\geq k$, it holds that
            $$\Delta\left(\left(f,f(X)\right), (f, U_m)\right)\leq \epsilon ,$$
            where $f\pgets \mathcal{F}$ distributed uniformly.
        \end{lemma}

        For each $z\in [l]$, $\langle H_z^{ij}, Y^{ij} \rangle$ is the $(i,j)$-th entry of $\langle H_z, Y \rangle_t$ by definition. Define a hash function family encoded by ${\bf H}_{\rm R}^{ij}$:
        $$F({\bf H}_{\rm R}^{ij}, Y^{ij})=({\rm vec}(H_1^{ij})~|\dots|~{\rm vec}(H_l^{ij}))^{\rm T}\cdot {\rm vec}({Y^{ij}})=\langle {\bf H}_{\rm R}^{ij}, Y^{ij} \rangle_F,$$
        Given that $l< \left(\frac{n}{t}\right)^2-\omega(\log n)$, and  $F$ is a pairwise independent hash function with input $Y^{ij}$, and random seed ${\bf H}_{\rm R}^{ij}$. By Leftover Hash lemma~(\cref{lmm:lhl}):
        \begin{align*}
            \Delta\left( \left({\bf H}_{\rm R}^{ij},\;\langle {\bf H}_{\rm R}^{ij}, Y^{ij} \rangle_F\right),\left({\bf H}_{\rm R}^{ij}, U_{\Fq^{l}}\right)\right) &=  \Delta\left( \left({\bf H}_{\rm R}^{ij},F({\bf H}_{\rm R}^{ij},Y^{ij})\right),
            \left({\bf H}_{\rm R}^{ij}, U_{\Fq^{l}}\right)\right)\\
            &\leq 2^{-\Omega ((n/t)^2-l)}\\
            &\leq \negl(n).
        \end{align*}
        
        Thus we have:
        \begin{align*}
            \Delta\left( \left({\bf H}_{\rm R},\langle {\bf H}_{\rm R}, Y\rangle_t\right),({\bf H}_{\rm R}, U_{(\Fq^{t^2})^l}) \right) &\leq t^2\cdot \Delta\left(({\bf H}^{ij}_{\rm R},\langle {\bf H}_{\rm R}, Y^{ij}\rangle_F),\; ({\bf H}_{}^{ij},U_{\Fq^{l}}) \right)\\
            &\leq \negl(n).
        \end{align*}
        Following the triangle inequality, we have
        $$\Delta\left( \left({\bf H}_{\bf A},\langle {\bf H}_{\bf A}, Y\rangle_t\right),\;\left({\bf H}_{\rm R},U_{\Fq^{t^2\times l}}\right)\right)\leq \negl(n),$$
        where ${\bf H}_{\bf A}\pgets ({\bf A}^{\perp_t})^l, {\bf H}_{\rm R}\pgets (\Fq^{n\times n})^l$, and $U_{\Fq^{t^2\times l}}$ follows uniform distribution.

        \medskip\medskip
        Thus any distinguisher $\algB$ for ${\rm dualMinRk}_{n,k,r}$ can be transformed into a distinguisher $\algA$ for ${\rm MinRk}({n,k,r})$ that runs the above reduction on input instance $({\bf H_A}, \langle {\bf H_A}, Y\rangle_t) \leftarrow \algR({\bf A},Y)$ and runs $\algB$ on $({\bf H_A}, \langle {\bf H_A}, Y\rangle_t) $. The resulting algorithm has an additional running time of 
        $\poly(n)$, and a negligible loss in advantage.
\end{proofof}

We finish with the proof of \cref{clm:subspace to dual space} stated above. 

\begin{proofof}{\Cref{clm:subspace to dual space}}
    For each matrix  $H_z \in \bf H_A$ (respectively for matrices in ${\bf H}_{\rm R}$), divide $H_z$ into $t^2$ submatrices as in the proof of \cref{lmm:reduce MinRank to dualMinRank}. Each submatrix is sampled identically and independently in both ${\bf H_A}$ and ${\bf H}_{\rm R}$, and thus we fix any entry $(i,j)\in [t]\times [t]$ and bound the distance $\Delta\left({\bf H}_{\bf A}^{ij}, {\bf H}^{ij}_{\rm R}\right)=\Delta\left({\rm vec}({\bf H}_{\bf A}^{ij}), {\rm vec}({\bf H}^{ij}_{\rm R})\right)$. Consider the following hybrids:
    \begin{itemize}
        \item $Hyb_1:$ 
        \begin{itemize}
            \item Sample $a_1,\dots,a_k \pgets \Fq^{(n/t)^2}$, and let ${\bf a}=(a_1,\dots,a_k )$. Let ${\bf a}^{\perp}=\{h\in \Fq^{(n/t)^2}|~\langle a_p, h \rangle=0\ \text{for}\ p \in [k] \}$
            \item Sample $h_1,\dots, h_l \pgets {\bf a}^{\perp}$ uniformly, and let ${\bf h}= (h_1,\dots, h_l)$.
            \item Output $({\bf a},{\bf h})$.
        \end{itemize}
        \item $Hyb_2:$ 
        \begin{itemize}
            \item Sample $a_1,\dots,a_k \pgets \Fq^{(n/t)^2}$ subject to $a_1,\dots, a_k$ being linearly independent, and let ${\bf a}=(a_1,\dots,a_k )$.
            \item Sample $h_1,\dots, h_l \pgets {\bf a}^{\perp}$ uniformly, and let ${\bf h}= (h_1,\dots, h_l)$.
            \item Output $({\bf a},{\bf h})$.
        \end{itemize}
        \item $Hyb_3:$ 
        \begin{itemize}
            \item Sample $a_1,\dots,a_k \pgets \Fq^{(n/t)^2}$ subject to $a_1,\dots, a_k$ being linearly independent, and let ${\bf a}=(a_1,\dots,a_k )$.
            \item Sample $h_1,\dots, h_l \pgets {\bf a}^{\perp}$ subject to $h_1,\dots, h_l$ being linearly independent, and let ${\bf h}= (h_1,\dots, h_l)$.
            \item Output $({\bf a},{\bf h})$.
        \end{itemize}
        \item $Hyb_4:$ 
        \begin{itemize}
            \item Sample $h_1,\dots,h_l \pgets \Fq^{(n/t)^2}$ subject to $h_1,\dots, h_l$ being linearly independent, and let ${\bf h}=(h_1,\dots,h_l )$.
            \item Sample $a_1,\dots, a_k \pgets {\bf h}^{\perp}$ subject to $a_1,\dots, a_k$ being linearly independent, and let ${\bf a}= (a_1,\dots, a_k)$.
            \item Output $({\bf a},{\bf h})$.
        \end{itemize}
                \item $Hyb_5:$ 
        \begin{itemize}
            \item Sample $h_1,\dots,h_l \pgets \Fq^{(n/t)^2}$, and let ${\bf h}=(h_1,\dots,h_l )$.
              \item Sample $a_1,\dots, a_k \pgets {\bf h}^{\perp}$ subject to $a_1,\dots, a_k$ being linearly independent, and let ${\bf a}= (a_1,\dots, a_k)$.
            \item Output $({\bf a},{\bf h})$.
        \end{itemize}
    \end{itemize}
    Here $Hyb_2$ samples ${\bf a}$ under the constraint that $a_1,\dots,a_k$ are linearly independent, similarly $Hyb_3$ enforces ${\bf h}$ to be linearly independent as well. $Hyb_4$ has the same distribution as $Hyb_3$, with a different sampling order. $Hyb_5$ samples ${\bf a}$ uniformly from ${\bf h}^{\perp}$.
    \begin{fact}[See, e.g., {\cite[Fact 2.2]{KRRSV20}}]\label{fct:distance fact}
    For any two distributions $X$ and $Y$, and event $E$,
    $$\Delta(X,Y)\leq \Delta(X|_E, Y)+\Pr_{X}[\neg E].$$
    \end{fact}
    
    Combining \Cref{fct:random matrix rank} and \Cref{fct:distance fact}, we have:
    $$\begin{array}{l}
    \Delta(Hyb_1,Hyb_2)\leq \q^{k-(n/t)^2} \quad \Delta(Hyb_2,Hyb_3)\leq \q^{k+l-(n/t)^2} \\
    \hspace{8em} \Delta(Hyb_4,Hyb_5)\leq \q^{l-(n/t)^2}
    \end{array}$$
    
    Given $(\frac{n}{t})^2-k-l=\omega(\log n)$, we derive using the data processing inequality that $$\Delta\left({\rm vec}({\bf H}_{\bf A}^{ij}), {\rm vec}({\bf H}^{ij}_{\rm R})\right)\leq \Delta(Hyb_1,Hyb_5)\leq \negl(n).$$ 
    Thus the distance between ${\bf H_A}$ and ${\bf H}_{\rm R}$ is bounded by:
    $$\Delta({\bf H_A},{\bf H}_{\rm R})\leq t^2 \cdot \Delta\left({\rm vec}({\bf H}_{\bf A}^{ij}), {\rm vec}({\bf H}^{ij}_{\rm R})\right)\leq \negl(n).$$
\end{proofof}

\input{search-to-decision-2.tex}

%% file: search-to-decision-2.tex
\subsection{Search-to-Decision Reduction}
\label{sec:s-to-d}

Here we will show that the decisional version of generic random MinRank (\Cref{dfn:minrank dist}) that we base our encryption scheme on, is equivalent in hardness to the search counterpart (\Cref{dfn:minrank-planted-search}). We will show this by means of a {\em search to decision} reduction, i.e., we will show how to convert any possible efficient distinguisher for decisional ge MinRank to an attacker for search MinRank with a comparable advantage, and a polynomial overhead in runtime. To do this, we will follow a fairly well-known technique for search-to-decision reductions first shown in \cite{ImpagliazzoN89} (who used it to show a similar result for the Subset Sum problem). Our result is as follows. 

\begin{theorem}[Search-to-Decision Reduction]\label{thm:StoDreduc}
For any polynomially bounded $k = k(n)$ and $r = r(n)$, suppose there exists a distinguisher $\algB$ for the Decision MinRank problem ${\rm MinRk}_{n,k,r}$ with advantage $\beta(n)$. Then, there is an algorithm $\algB'$ for the Search MinRank problem ${\rm SearchMinRk}_{n,k,r}$ that succeeds with probability $\Omega(\beta(n)^3)$. Further, if the runtime of $\algB$ is bounded by $T_{\algB}(n)$, then that of $\algB'$ is bounded by $T_{\algB}(n)\cdot\poly(k,\log 1/\beta(n))$. 
\end{theorem}

\begin{corollary}
    \label{cor:StoDreduc}
    For any set of polynomially bounded parameters, if there is a polynomial-time algorithm for the Decision MinRank problem that has non-negligible advantage, then there is a polynomial-time algorithm for the Search MinRank problem with the same parameters that has non-negligible success probability.
\end{corollary}

\begin{remark}
    We focus on MinRank on $\mathbb{F}_\q$ in this work for simplicity, but our search-to-decision reduction and its proof apply to all $\mathbb{F}_q$ of any order $2\leq q\leq \poly(n)$ straightforwardly.
\end{remark}

\begin{proofof}{\cref{thm:StoDreduc}}
    Suppose, without loss of generality, there exists an adversary $\algB$ that accepts with probability $\beta(n)$ more on an input from $\mathcal{D}_{{\rm MinRk}(n,k,r)}$ than on input that is uniform over $\left(\mathbb{F}_\q^{n\times n}\right)^{k+1}$. We proceed to construct a predictor takes as input a vector $x\in \mathbb{F}_{\q}^k$ and $({\bf A}, Y)\pgets \mathcal{D}_{{\rm MinRk}(n,k,r)}$, with $Y={\bf A}(s)+E$ for some private secret $s$, and predicts the  inner product $\langle s,x\rangle$
    with non-trivial advantage. Looking ahead, we combine the predictor with the Goldreich-Levin theorem to complete the proof.

    \begin{lemma}[Goldreich-Levin~\cite{AGS03,GoldreichL89}]\label{lmm:gl} 
        For any $\epsilon\in (0,1)$, there exists an explicit algorithm $\algA_{\rm GL}$ parameterized by $n$ and $\eps$ and having oracle access to an algorithm $\algA$ that takes input a vector $x\in \mathbb{F}_\q^k$ and outputs a value $b\in \mathbb{F}_\q\cup \{\perp\}$, satisfying the following. For any $s\in\mathbb{F}_\q^k$, if $\algA$ is such that $\Pr\limits_{x\pgets \mathbb{F}_{\q}^k}\left[\algA(x)=\langle x,s\rangle\right]\geq 1/\q + \epsilon$, then $\algA_{\rm GL}$ outputs $s$ with probability at least $\varepsilon^2$. If $\algA$ runs in time $T$, then $\algA_{\rm GL}$ runs in time $\poly(k, \log (1/\varepsilon))\cdot T$.
    \end{lemma}
    We construct the predictor as follows:
  \begin{framed}
\underline{Algorithm $\mathsf{Pred}^\algB(({\bf A},Y),x)$:}
\begin{enumerate}
    \item Parse ${\bf A}=(A_1,\dots,A_k)$ and $x=(x_1,\dots, x_k)$.
    \item Sample a uniform random $b\pgets \Fq$
        \item Sample $M \gets \mathbb{F}_\q^{n \times n}$ uniformly.
        \item Set $A_i' \gets A_i + x_i\,M$ for all $i\in[k]$, let ${\bf A}'=(A_1',\dots,A_k')$, and set $Y' \gets Y + b\,M$.
        \item Query $\algB$ on $({\bf A}',Y')$. If it accepts, output $b$; otherwise, output $1-b$.
\end{enumerate}
\end{framed}

\noindent For any planted instance $({\bf A},E,s)$, query $x$, matrix $M$, and $b\in\Fq$, we have the following:
$$
{\bf A}'(s)=\sum_i s_i(A_i+x_iM)={\bf A}(s)+\langle s,x\rangle M,\qquad
Y' = {\bf A}'(s)+E+(b-\langle s,x\rangle)M.
$$
If we condition on $b=\langle s,x\rangle$, then $({\bf A}',Y')\sim\mathcal{D}_{{\rm MinRk}}$ for randomly sampled $({\bf A},E,s)$ and any $M$. On the other hand, if we condition on $b \neq \langle s,x\rangle$, then for random $({\bf A},E,s,M)$, the $({\bf A}',Y')$ is also uniformly random. Note that the event $b = \langle s,x\rangle$ happens with probability $1/2$ in the algorithm. Thus, the probability that ${\sf Pred}^{\algB}$ outputs $\langle s,x\rangle$ is:

\begin{align*}
&\Pr\limits_{{\bf A}, s, E, x, M,\algB}\left[{\sf Pred}^{\algB}(({\bf A},{\bf A}(s)+E),x)=\langle s,x\rangle\right]\\
&= \pr{b = \ip{s,x}}\cdot \pr{\algB({\bf A}',Y')=1\ |\ b=\ip{s,x}} \\ & \hspace{2em} + \pr{b \neq \ip{s,x}}\cdot \pr{\algB({\bf A}',Y')=0\ |\ b\neq\ip{s,x}}\\
&= \frac{1}{2} \cdot \prob{({\bf A}',Y')\gets \mathcal{D}_{{\rm MinRk}}}{\algB({\bf A}',Y')=1} \\ & \hspace{2em}+ \frac{1}{2}\cdot \left(1- \prob{({\bf A}',Y')\gets (\Fq^{n\times n})^{k+1} }{\algB(({\bf A}',Y'))=1}\right)\\
&= \frac{1}{2} + \frac{\beta(n)}{2}
\end{align*}

For any  ${\bf A}, s, E$, define the quantity 
\iflncs
\else
$\Delta({\bf A},s,E)= \Pr_{x,M,\algB}\left[{\sf Pred}^{\algB}(({\bf A},{\bf A}(s)+E),x)=\langle s,x\rangle\right]-\frac{1}{2}$,
\fi
 and  we have 
$\Delta({\bf A},s,E)\leq \frac{1}{2}$ and $\underset{{\bf A}, s, E}{\mathbb{E}}\left[\Delta({\bf A}, s, E)\right]=\frac{\beta(n)}{2}$. Then there is at least $\frac{\beta(n)}{4}$ fraction of $({\bf A}, s, E)$ that satisfies $\Delta({\bf A}, s, E)\geq \frac{\beta(n)}{4}$; otherwise for any $\zeta<\frac{\beta(n)}{4}$, and if there is only a $\zeta$ fraction of $({\bf A}, s,E)$ with $ \Delta({\bf A}, s, E)\geq \frac{\beta(n)}{4}$, then the  $\underset{{\bf A},s,E}{\mathbb{E}}\left[\Delta\left({\bf A},s,E\right)\right]\leq\zeta\cdot \frac{1}{2}+(1-\zeta)\cdot(\frac{\beta}{4})<\frac{\beta(n)}{2}$, a contradiction.

Thus, for at least a $\beta(n)/4$ fraction of $({\bf A},s,E)$, the probability over $x,M$ and $\algB$ that ${\sf Pred}^{\algB}(({\bf A},{\bf A}(s)+E),x)$ outputs $\ip{s,x}$ is at least $1/2 + \beta(n)/4$. Whenever such an instance is sampled, the algorithm guaranteed by \cref{lmm:gl} successfully recovers $s$ with probability at least $\Omega(\beta(n)^2)$. So overall, the search algorithm succeeds with probability at least $\Omega(\beta(n)^3)$. The running time follows from observing that ${\sf Pred}^{\algB}$ has almost no overhead over $\algB$, and from the runtime bound in \cref{lmm:gl}.

\end{proofof}

%% file: pke.tex
\section{PKE from MinRank}
\label{sec:pke}

In this section, we present our Public-Key Encryption scheme and prove its correctness and security assuming hardness of MinRank problem with suitably chosen parameters. Our scheme ${\sf PKE}=({\sf KeyGen}, {\sf Enc}, {\sf Dec})$ is parameterized by the security parameter $n$ and polynomially bounded functions $k(n)$, $r(n)$ and $t(n)$, and is presented in \Cref{con:pke}. The following theorem captures its properties.

\begin{figure}[!ht]
\input{figures/pke-construction}
\caption{PKE from MinRank}\label{con:pke}
\end{figure}

\begin{samepage} 
\begin{theorem}\label{thm:pke from minrank}
    Consider polynomially bounded $k=k(n)$, $t=t(n)$ and $r=r(n)$ satisfying the following conditions:
    \begin{enumerate}
        \item $r(n)^2 < t(n)-\log(n)$, $(n/t)^2-2k-1=\omega(\log n)$, and $t$ divides $n$
        \item The decision MinRank problem ${\rm MinRk}_{n,k,r}$ is hard for all polynomial-time algorithms
    \end{enumerate}
    When instantiated with these parameters, the scheme presented in \cref{con:pke} is a public-key encryption scheme achieving semantic security and correctness. 
    Further, if there is an algorithm running in time $T(n)$ that breaks the security of this scheme, then there is an algorithm running in time $T(n) + \poly(n)$ that has non-negligible advantage in solving ${\rm MinRk}_{n,k,r}$.
\end{theorem}
\end{samepage}

\paragraph{Choices of parameters.} To fully specify the PKE scheme, we need to specify the parameters $k$, $r$, and $t$ as functions of the security parameter $n$. These need to be chosen in such a way that the scheme is correct, the duality of MinRank holds, and the MinRank problem is hard. The requirements for the first two are already specified in the hypothesis of \cref{thm:pke from minrank}: $t-r^2>\log n$ and $(n/t)^2-2k= \omega(\log n)$, respectively. We ignore for now the requirement that $t$ has to divide $n$, as this may be arranged with small perturbations of their values.

As shown in \cref{thm:pke from minrank}, the complexity of breaking the security of the scheme given a single ciphertext is tightly related to the complexity of solving MinRank. By standard hybrid arguments, a similar relation continues to hold even given multiple ciphertexts, except in some extreme cases. For this reason, we will use the complexity of solving MinRank itself as a proxy for the security of our scheme in our discussion here. Thus, in order to select secure parameters, we need to take into account the best known algorithms for the MinRank problem. A summary of these algorithms is presented in \Cref{table:attacks}, with detailed discussion in the rest of \Cref{sec:attacks}.

As is apparent from \cref{table:attacks}, the various algorithms for MinRank have different requirements of and dependencies on the parameters $n$, $k$, and $r$. This leads to many meaningful settings for these parameters, each with different tradeoffs between security and efficiency. In general, increasing $k$ and $r$ makes the algorithms less efficient, but these are also bounded by the conditions above be at most roughly $(n/t)^2$ and $\sqrt{t}$, respectively. The following are a few natural settings. The constants below are to be chosen so that the conditions for correctness and duality are met.

\begin{enumerate}
    \item $t = \Theta(n^{1/2})$, $k = \Theta(n)$, $r = \Theta(n^{1/4})$.

    In this case, the most efficient algorithm is the Kernel attack, which runs in time $\approx 2^{r\cdot \lceil k/n \rceil} = 2^{O(n^{1/4})}$. The public keys and ciphertexts are of size roughly $n^3$ and $n^2$, respectively.

    \item $t = \Theta(\sqrt{\frac{n}{\log n}})$, $k = \Theta(n\log n)$, $r = \Theta\left((\frac{n}{\log n})^{1/4}\right)$.
    
    In this case, the most efficient algorithms are the Kernel attack and Support Minor attack (with linearization), both of which run in time $2^{O(n^{1/4}\cdot \log^{3/4} n)}$, up to polynomial factors. The public keys and ciphertexts are of size roughly $n^{3}\log{n}$ and $n^2/\log{n}$, respectively. This is the setting of parameters that leads to the largest running times of the algorithms listed as a function of $n$, according to the complexity estimates in \cref{table:attacks}.
    
    
    \item $t = \Theta(\log^6{n})$, $k = \Theta\left(\frac{n}{\log n}\right)$, $r = \Theta(\log^3{n})$.

    In this case, the most efficient algorithm is the Kipnis-Shamir attack (with XL), which runs in time $\approx 2^{O(\log^2{n}\log\log{n})}$. The public keys and ciphertexts are of size roughly $(n^3/\log{n})$ and $(n\log^{11}{n})$, respectively.
\end{enumerate}

\subsection{Proof of Correctness and Security}
\label{sec:pke-proof}

\begin{proofof}{\Cref{thm:pke from minrank}}
We first show correctness of the scheme. 

\paragraph{Correctness.} In case $x=0$, ${\bf ct}=\langle R, {\bf A'}\rangle_t$, where ${\bf A'}=({\bf A},\; {\bf A}(s)+E)$. Thus,
\begin{align*}
  M 
  &=\langle R,\; {\bf A}(s)+E \rangle_t - \sum\limits_{i\in [k]}\; s_k\cdot \langle R, A_i\rangle_t\\
  &= \langle R, {\bf A}(s)+E-{\bf A}(s)\rangle_t\\
  &= \langle R, E\rangle_t.
\end{align*}

Thus, ${\sf rank}(M)= {\sf rank}(\langle E, R\rangle_t)\leq r^2< t-\log n$.  

On the other hand if $x=1$, $C=(C_1,\dots,C_k,C_{k+1})$ where $C_1,\dots, C_k, C_{k+1} \pgets \mathbb{F}_\q^{t\times t}$ are independently random matrices. Then  $M=C_{k+1}-\sum_{i} s_i\cdot C_i$ is a random $t\times t$ matrix, which, by \Cref{fct:random matrix rank}, has rank greater than $t-O(\log^{2/3} n)$ with $1-\negl(n)$ probability.

\paragraph{Semantic Security.} We show semantic security by considering the following hybrids:


\begin{framed}
\noindent $Hyb_1(1^n)$:
\begin{enumerate}
  \item $(sk,pk) \leftarrow {\sf KeyGen}(1^n)$ with $pk = \bigl({\bf A},\,Y = {\bf A}(s) + E\bigr)$, where 
  \[
    \quad  {\bf A} \pgets (\Fq^{n\times n})^k,
    \; s \pgets \Fq^{k}  \quad  E\pgets \Fq^{n\times n} \text{ s.t. }{\sf rank}(E)\leq r. 
  \]
  \item $ct\leftarrow {\sf Enc}( pk, 0)$ with  $ct = \bigl(\langle R, {\bf A}\rangle_t,\;\langle R,Y\rangle_t\bigr),$ where \[
     R \pgets \Fq^{n\times n}\text{ s.t. } {\bf rank}(R)\leq r.
  \]
  \item Output $(pk,ct)=\left({\bf A}, Y, \langle R, {\bf A}\rangle_t, \langle R, Y\rangle_t\right)$.
\end{enumerate}
\end{framed}
\begin{framed}
\noindent $Hyb_2(1^n)$:
\begin{enumerate}
  \item Sample $A_1,\dots, A_k, Y\pgets \Fq^{n\times n}$, and set ${\bf A}=(A_1,\dots,A_{k})$.
  \item Sample $R \pgets \Fq^{n\times n}\text{ s.t. } {\bf rank}(R)\leq r$.
  \item Output $({\bf A},Y,\langle R, {\bf A}\rangle_t, \langle R, Y\rangle_t)$.
\end{enumerate}
\end{framed}
\begin{framed}
\noindent $Hyb_3(1^n)$:
\begin{enumerate}
  \item Sample $A_1,\dots, A_k, A_{k+1}\pgets \Fq^{n\times n}$, and set ${\bf A}'=(A_1,\dots,A_{k},A_{k+1})$.
  \item Sample $R \pgets \Fq^{n\times n}\text{ s.t. } {\bf rank}(R)\leq r$.
  \item Output $({\bf A}',\langle R, {\bf A}'\rangle_t)$.
\end{enumerate}
\end{framed}
\begin{framed}
\noindent $Hyb_4(1^n)$:
\begin{enumerate}
  \item Sample $A_1,\dots, A_k, A_{k+1}\pgets \Fq^{n\times n}$, and set ${\bf A}'=(A_1,\dots,A_{k},A_{k+1})$.
  \item Sample $C_1,\dots,C_{k}, C_{k+1}\pgets \Fq^{t\times t}$, and set ${\bf C'}=(C_1,\dots,C_{k}, C_{k+1})$.
  \item Output $({\bf A}',{\bf C}')$.
\end{enumerate}
\end{framed}
\begin{framed}
\noindent $Hyb_5(1^n)$:
\begin{enumerate}
  \item $(sk,pk) \leftarrow {\sf KeyGen}(1^n)$ with $pk = \bigl({\bf A},\,Y = {\bf A}(s) + E\bigr)$, where 
  \[
    \quad  {\bf A} \pgets (\Fq^{n\times n})^k,
    \; s \pgets \Fq^{k}  \quad  E\pgets \Fq^{n\times n} \text{ s.t. }{\sf rank}(E)\leq r. 
  \]
  \item $ct\leftarrow {\sf Enc}( pk, 1)$ with  $ct = {\bf C'}=(C_1,\dots, C_{k}, C_{k+1})$.
  
  \item Output $(pk,ct)=\left(({\bf A}, Y), {\bf C'}\right)$.
\end{enumerate}
\end{framed}
Note that $Hyb_1(1^n)$ corresponds to an encryption of $0$ and $Hyb_5(1^n)$ corresponds to an encryption of $1$. The $Hyb_2(1^n)$ (resp. $Hyb_4(1^n)$) replaces $Y={\bf A}(s)+E$ in $Hyb_1(1^n)$ (resp. in $Hyb_5(1^n)$) with independently uniform matrix $Y \pgets \Fq^{n\times n}$ (resp. $A_{k+1} \pgets \Fq^{n\times n}$). By the assumed hardness of ${\rm MinRk}_{n,k,r}$, we have the following. 
\begin{claim}
    $Hyb_1(1^n) \approx_c Hyb_2(1^n)$ and $Hyb_4(1^n) \approx_c Hyb_5(1^n)$.
\end{claim}
The $Hyb_3(1^n)$ is the same distribution as $Hyb_2(1^n)$ by renaming $A_{k+1}=Y$ and setting ${\bf A}'=({\bf A}, A_{k+1})$. Next, note that the distributions in $Hyb_3$ and $Hyb_4$ are exactly those that appear in the definition of the dual MinRank problem ${\rm dualMinRk}_{n,k+1,r}$. Applying the duality lemma (\Cref{lmm:reduce MinRank to dualMinRank}) with $l=k+1$, and the assumed hardness of ${\rm MinRk}_{n,k,r}$, we also have the following.
\begin{claim}
    $Hyb_3(1^n) \approx_c Hyb_4(1^n)$.
\end{claim}
Combining the above claims shows semantic security. 

\paragraph{Tight reduction.} To prove the last part of the theorem, suppose there is an algorithm that runs in time $T(n)$ and distinguishes between $Hyb_1(1^n)$ and $Hyb_5(1^n)$ with non-negligible advantage, then there should be a consecutive pair of hybrids that it distinguishes with non-negligible advantage as well. Distinguishing between each pair of consecutive hybrids corresponds exactly to solving either the ${\rm MinRk}_{n,k,r}$ or the ${\rm dualMinRk}_{n,k+1,r}$ problem with the same advantage. Using \cref{lmm:reduce MinRank to dualMinRank} and standard arguments, this implies that there is also an algorithm that runs in time $T(n)+\poly(n)$ and has non-negligible advantage in solving ${\rm MinRk}_{n,k,r}$.
\end{proofof}

%% file: figures/pke-construction.tex
\begin{framed}
\noindent \textbf{\underline{PKE from MinRank}}\\


\noindent{\sf Parameters:} $k=k(n)$, $r=r(n)$ and $t=t(n)$.

\vspace{1em}
\noindent$\underline {{\sf KeyGen}(1^n)}$:
    \begin{enumerate}
        \item Sample $\td\pgets \mathbb{F}_2^k;\; {\bf A}\pgets (\mathbb{F}_2^{n\times n})^k ;\; E\pgets \mathbb{F}_2^{n\times n}\text{ s.t. }{\rm rank}(E)\leq r$.
        \item Set $sk=\td $ and $pk=({\bf A}, {\bf A}(\td)+E)$.
    \end{enumerate}

\vspace{1em}
\noindent $\underline {{\sf Enc}( pk, x\in \{0,1\})}$: 
    \begin{enumerate}
        \item Parse $pk=(A_1',A_2',\dots, A'_{k+1})={\bf A'}$.
        \item If $x=1$, sample $k+1$ random matrices $(V_1,\dots,V_{k+1})\leftarrow (\mathbb{F}_2^{ t \times t})^{k+1}$; and set these to be the ciphertext $ct$.
        \item Else if $x=0$, sample a random matrix $R\leftarrow \mathbb{F}_2^{n\times n}$ under the constraint that ${\rm rank}(R)\leq r$; set the ciphertext to be ${\bf ct}=\langle R, {\bf A'}\rangle_t$.
        \item  The encryption algorithm outputs $ct$. 
    \end{enumerate}
    
\vspace{1em}
\noindent $\underline {{\sf Dec}(sk=\td, ct)}$: 
    \begin{enumerate}
        \item Parse ${\bf ct}=(C_1,C_2,\cdots, C_k, C_{k+1})$.
        \item Set $ M=C_{k+1}-\sum\limits_{i\in [k]}s_i\cdot C_i$.
        \item If  ${\rm rank} (M)<t-\log^{2/3}n$ output $0$; otherwise output $1$. 
    \end{enumerate}
\end{framed}

%% file: analysis.tex
\section{Algorithms for MinRank} \label{sec:attacks}\label{sec:security analysis}

Herein we focus on attacks on the square-matrix MinRank problem since this setting is the most relevant to our cryptosystem, although most of these attacks we describe generalize (usually quite naturally) to the general rectangular matrix case. These attacks can be broadly categorized into algebraic and combinatorial approaches.  
A summary of these attacks, their estimated complexities, and where they appear is given in \cref{table:attacks}. These estimates are for the settings of the parameters in the algorithms that enable them to succeed with at least some constant probability. \cref{table:attackLBs} lists lower bounds on the complexity of some of these attacks that follow from elementary considerations like the number of variables, size of the equation system they construct, etc..

\begin{table}[h!]
\centering
\input{figures/attack-overview}
\caption{Complexity estimates of prominent MinRank algorithms} \label{table:attacks} 
\end{table}


\begin{remark}
Except for brute-force search, all algorithms discussed in this section are \emph{heuristic}, and their stated complexities are \emph{average-case}, typically valid only for generic random instances drawn from $\mathcal{D}_{\mathrm{MinRk}(n,k,r)}$. These bounds are not worst-case guaranties, and adversarially structured instances can be constructed on which these algorithms behave worse than the stated estimates.
\end{remark}

\begin{table}[!ht]
\centering
\input{figures/simpleLBs}
\caption{Simple lower bounds on complexity of the prominent attacks} \label{table:attackLBs} 
\end{table}

\subsection{Combinatorial Attacks}

\paragraph{Additional Notation.}
For a matrix $A\in \mathbb{F}_\q^{n\times m}$, let ${\rm Ker}(A)=\{v\in \mathbb{F}_\q^m\;|\; Av=0\}$ denote the right kernel space of $A$.
\subsubsection{Brute force Attack} \label{subsubsec:brute force}

The simplest attack involves trying all possible linear combinations $s \in \mathbb{F}_\q^k$ and testing whether any of them hit the target rank $r$, and its complexity is $O(2^k\cdot n^{\omega})$, where $\omega$ is the matrix multiplication exponent.

\begin{samepage}
Alternatively, one can instead enumerate all matrices $E$ of ${\rm rank}(E)\leq r$ and test if $(Y-E)$ lies in the span of ${\bf A}=(A_1,A_2,\dots, A_k)$. There are $2^{2nr-r^2-r}\leq \sum\limits_{i=1}^r \frac{\left(\prod\limits_{i=0}^{r-1}\; (\q^n-\q^i)\right)^2}{\prod\limits_{i=0}^{r-1}\;(\q^r-\q^i)}\leq r\cdot \q^{2nr-r^2+r}$ matrices of rank up to $r$. This thus takes time $O\left( \q^{2nr-r^2+r}\cdot r\cdot  n^{2\omega}\right)$.
\end{samepage}

\subsubsection{Kernel Attack}
The Kernel Attack was introduced by Goubin and Courtois \cite{goubinCryptanalysisTTMCryptosystem2000}. The key observation here is that if $r$ is relatively small, any desired combination of $E=Y-A(s)=Y-\sum_{i=1}^{k} s_i A_i$ of rank $r$ will have a substantially large kernel (of dimension at least $n-r$). In particular, if one uniformly samples a vector of length $n$, this will be in the kernel of $E=Y-A(s)$ with probability at least $2^{-r}$. 

Sample $\lceil k/n \rceil$ linearly independent  vectors $y_1,y_2,\dots,y_{[\lceil k/n \rceil ]}$ that are hopefully in the kernel of the unknown resulting matrix $E$. Use these vectors to construct a system of linear equations over the $s_i$'s, and attempt to solve for the latter (using Gaussian elimination).

Observe that each $y_j$ potentially yields $n$ linear equations in the $s_i$'s, corresponding to $E\cdot y_j = 0^n$. Now if all of the vectors $y_j$ indeed fall inside the kernel of $E$, for a random MinRank instance as defined in \Cref{dfn:minrank-planted-search}, the system of linear equations is overdetermined with high probability 
(since we have $n\cdot\lceil k/n\rceil$ linear equations in $k$ variables, and with high probability over the sampling of instance and $y_i$'s, most of them are linearly independent). The probability that they all fall in the kernel if they are sampled uniformly at random is $2^{-r\cdot \ceil{k/n}}$. Thus the expected running time of the algorithm on random MinRank instances is $O(2^{r\cdot \lceil k/n \rceil}\cdot k^{\omega})$, where $\omega$ is the matrix multiplication exponent.

\subsection{Algebraic Attacks}


The algebraic attack methods (e.g. Kipnis-Shamir \cite{KS99}, minors attack \cite{FS90}, support minors \cite{bardetImprovementsAlgebraicAttacks2020a}) share a similar paradigm : (i) Given a MinRank instance and a target rank $r$, construct an algebraic model that generates a  polynomial system over $\Fq$, whose solution are in one-to-one correspondence with solutions to the given MinRank instance; (ii) solve the polynomial system using a state-of-the-art algorithm for solving multivariate polynomial system of equations, such as \groebner Basis methods \cite{buchbergerBrunoBuchbergersPhD2006,Fau02,Fau99} or the XL algorithm \cite{CKPS00}.

We will present these algebraic attacks in this subsection, and discuss the overall complexity when combined with polynomial system solvers.

\subsubsection{Kipnis-Shamir Attack}

Kipnis and Shamir's cryptanalysis of the HFE cryptosystem \cite{KS99} gives rise to an attack on the MinRank problem. Essentially, the attack recasts MinRank as solving a system of multivariate quadratic (MQ) equations. 

Let $A_1,\dots, A_k, Y\in \mathbb{F}_\q^{n\times n}$ be the MinRank instance, and set 
$E=E(s)=Y-\sum\limits_{i=1}^{k} s_i\cdot A_i.$ 
If $E$ is a random matrix such that ${\rm rank}\left(E\right)\leq r$, then with good probability, the first $r$ columns of $E$ are independent while all the other columns can be written as linear combinations of them. If this happens, $E$ has $(n-r)$ linearly kernel vectors of the form of the columns of the following matrix:
$$K=K(y)=\begin{pmatrix}
    1&0&\dots& 0\\
    0&1& \dots& 0\\
    \vdots& \vdots& \ddots& \vdots\\
    0& 0& \dots& 1\\
    y_1^{(1)}& y_1^{(2)}& \dots& y_1^{(n-r)}\\
    \vdots & \vdots& \ddots& \vdots\\
    y_r^{(1)}& y_r^{(2)}& \dots& y_r^{(n-r)}
\end{pmatrix}.$$
Kipnis-Shamir considers the equation $E(s)\cdot K(y)=0^{n\times (n-r)}$.
If $(s,y)$ is a solution to this equation, then $s=(s_1,\dots,s_{k})$ is a solution to the original MinRank instance. Thus, the attack just needs to solve this system of $n(n-r)$ bilinear equations in the $(k + r(n-r))$ variables $(s,y)$. This may be done using various methods.

\paragraph{Linearization.} Replacing each monomial of the form $s_i\cdot y_j^{(l)}$ with an independent variable yields a linear system with at most $\left(k \cdot r\cdot (n-r)\right)$ unknowns. Hence, when $n \gg k\cdot r$, the resulting linear system is, for random instances, overdetermined with high probability over the sampling of instances and can therefore be solved in polynomial time. \cite{KS99} further introduced the technique of {\em relinearization}, which solves the bilinear system with relatively with fewer equations and relaxes the condition to $n=\Omega( k\cdot r)$.

\paragraph{XL algorithm.} Multiply each bilinear equation by all monomials up to a chosen degree so that, after linearization, the resulting system becomes overdetermined if the chosen degree is large enough. Solving the final linear system dominates the cost and, for random instances, takes running time $O\left(\left(r\cdot(\frac{k(r+1)}{n+k})\right)^{\left(\frac{k(r+1)}{n+k}+1\right)\cdot \omega}\right)$. See e.g., {\cite[Section 5]{VBCPS19}}, for the analysis. 


\paragraph{\groebner Basis algorithms.} Compute a \groebner basis for the polynomial system (e.g., via F4~\cite{Fau99}, F5~\cite{Fau02}) and recover a solution from it. These algorithms bear similarities to the XL algorithm, and their complexity is often bounded in terms of the \emph{degree of regularity} of the given system. For random MinRank instances, the best upper bound on the degree of regularity is $d_{\rm reg}\leq \min\left((n-r)\cdot r,~k\right)+1$, and the complexity of the algorithms is $O\left(\binom{k+r(n-r)+d_{\rm reg}}{d_{\rm reg}}^{\omega}\right)$. See e.g., {\cite[Section 4.1]{FEDS10}} for the analysis.

\paragraph{Quantum algorithms.} 
This is described in \cite[Section 4.2.1]{AdjBBERSVZ24}. The idea here is to take the system of equations in the Kipnis-Shamir formalism, and rewrite this as a linear system with the elements in $K$ as the variables. This allows them to set up a quantum search problem. They get an overall attack running in time $O(\q^{r/2\cdot \lceil k/n\rceil})$.  

\subsubsection{Minors Attack }

The Minors attack, which some consider folklore, was first published in \cite{Courtois01b}, with rigorous analyses provided by \cite{FEDS10}, and further improved by \cite{faugereComplexityGeneralizedMinRank2013, {GND23}}. 
The attack is based on the observation that $s=(s_1,\dots,s_k)$ is a solution to a planted search MinRank instance $\left({\bf A}, Y\right)=\left((A_1,\dots,A_k),\; Y\right)$ if and only if all size-$(r+1)$ minors of the matrix $E=Y-\sum_{i} s_i\cdot A_i$ vanish. Recall that a minor of a matrix of size $(r+1)$ is the determinant of some $(r+1)\times (r+1)$-dimensional sub-matrix.


Consider the matrix $E(s) = Y-\sum_i s_i\cdot A_i$ whose entries are linear expressions in the variables $s_1,\dots,s_k$. There are $\binom{n}{r+1}^2$ minors of $E(s)$ of size $(r+1)$, and each of these is a degree-$(r+1)$ polynomial in the $s_i$'s. Requiring that all of these are $0$ then gives us a corresponding polynomial system in $k$ variables that then remains to be solved. Note that if the attack is to explicitly write down this system, this would already take at least $\binom{n}{r+1}^2$ time, which is super-polynomial in $n$ if $r$ is $\omega(1)$.

For random MinRank instances, \cite{FEDS10} showed that the degree of regularity of the above system of equations is bounded by $n(n-r)-k+1$, leading to a complexity of $O\left(\binom{n(n-r)+1}{k}^{\omega}\right)$ of solving it using \groebner Basis algorithms. If $e(k+r)\leq \frac{n^2}{(r+1)}$, there are  $\binom{n}{r+1}^2$  equations, which is more than than $\binom{r+k}{r+1}$ -the number of monomials of degree $r+1$. We can thus use linearization to solve the polynomial system.

\subsubsection{Support Minors Attack}

The Support Minors attack proposed in \cite{bardetImprovementsAlgebraicAttacks2020a} is an ``economic'' adaptation of the Minors attack. The attack uses the fact that if $E(s)$ has rank at most $r$, then there must exist a subspace of $\Fq^n$ of dimension $r$ such that the rows of $E(s)$ are contained in it. The objective is then to find a basis for such a subspace. We model this basis as the rows of an $r\times n$ matrix $C$ of new variables, and considers the matrix $D=\begin{bmatrix}E(s)\\C\end{bmatrix}$. We then require that all the size-$(r+1)$ minors of $D$ that involve exactly one row from the $E(s)$ part evaluate to $0$. Note that there are now only $n\cdot \binom{n}{r+1}$ such minors, as opposed to $\binom{n}{r+1}^2$ in the Minor attack. This comes at the cost of a larger number of variables -- $(k+rn)$ here, as opposed to $k$ in the Minor attack. Again, note that any attack that explicitly writes down the entire system takes at least $n\cdot \binom{n}{r+1}$ time.

If $n\cdot \frac{n-r}{r+1}\geq k$, we can use linearization to solve the polynomial system. Each variable in the linearization corresponds to the product of one of the $k$ variables $s_1,\dots,s_k$ and the determinant of an $r\times r$ sub-matrix of $C$. This results in $k\binom{n}{r}$ variables, which is less than the number of equations if the above condition is satisfied. Due to the constraints in \cref{thm:pke from minrank}, our setting of parameters always satisfies the above condition, and so this attack is always applicable. The efficiency of XL and hybrid algorithms for solving the above polynomial system have also been studied, and these can work even when the above condition is not satisfied~\cite{BG25,bardetRevisitingAlgebraicAttacks2022}.

%% file: figures/attack-overview.tex
\begingroup
\renewcommand{\arraystretch}{1.8}

\begin{tabular}{|l|c|c|c|}
\hline
\multicolumn{1}{|c|}{\textbf{Attack}} & \textbf{Complexity} & \textbf{Remarks} & \textbf{Ref.}\\
\hline
Brute Force Attack& $\min\left(2^k, 2^{2r(n-r)+r}\right)$  & --- & \Cref{subsubsec:brute force}\\ 
Kernel Attack & $2^{r\cdot \lceil k/n \rceil}$  & --- &\cite{goubinCryptanalysisTTMCryptosystem2000}\\ 
Kipnis--Shamir + Linearization & $\poly(n)$ & needs $n\geq \Omega(k\cdot r)$ & \cite{KS99}\\
Kipnis--Shamir + XL & $\left(r\cdot \frac{kr}{n+k} \right)^{\left(\frac{kr}{n+k}+1\right)\cdot \omega}$ & --- & {\cite{VBCPS19}} \\ 
Kipnis--Shamir + Gr\"obner & $\binom{k+r(n-r)+d}{d}^{\omega}$ & $d= \min\left(k,\; r(n-r)\right)$ & {\cite{FEDS10}} \\ 
Minors + Linearization & $\binom{n}{r}^{2\omega}$ & needs $\frac{n^2}{(r+1)} \geq \Omega(k+r)$ & \cite{bardetImprovementsAlgebraicAttacks2020a}\\
Minors + Gr\"obner& $\binom{n(n-r)+1}{k}^{\omega}$ & --- & \cite{FEDS10}\\ 
Support Minors + Linearization & $\left(k\cdot \binom{n}{r}\right)^\omega$ & needs $k\leq \frac{n(n-r)}{r+1}$ & \cite{bardetImprovementsAlgebraicAttacks2020a}\\
\hline
\end{tabular}
\begin{tablenotes}\footnotesize
\item $n$ - dimension of matrices, $k$ - number of matrices, $r$ - target rank, $\omega$ - matrix multiplication constant
\item For brevity, we have simplified some expressions and ignored multiplicative factors polynomial in $n$
\end{tablenotes}
\endgroup

%% file: figures/simpleLBs.tex




\begingroup
\renewcommand{\arraystretch}{1.8}

\begin{tabular}{|l|c|}
\hline
\multicolumn{1}{|c|}{\textbf{Attack}} & \textbf{Complexity Lower Bbounds} \\
\hline

Brute Force Attack& $\min\left(2^k, 2^{2rn-r^2-r}\right)$  \\
Kernel Attack & $\Omega\left(2^{r\cdot \lceil k/n \rceil}\right)$   \\
Kipnis-Shamir Approach & $\Omega\left(n(n-r)\right)$  \\
Minors Approach & $\Omega\left(\binom{n}{r+1}^2\right)$  \\
Support Minors Approach & $\Omega\left( n\cdot \binom{n}{r+1}\right)$   \\

\hline
\end{tabular}
\begin{tablenotes}\footnotesize
\item Parameters: $n$ - dimension of matrices, $k$ - number of matrices, $r$ - target rank
\item  The bound for the first two attacks are obtained directly from the complexity analysis; the lower bounds for the rest algebraic attacks are estimated based on the number of distinct equations arising in the system. 
\end{tablenotes}

\endgroup

%% file: history.tex
\section{A History of the MinRank problem}\label{sec:history}



The MinRank problem was introduced in \cite{BFS99} and shown to be NP-hard in the worst-case. 
The problem has subsequently been widely studied in the context of cryptography from multivariate equations, where it crops up regularly and thus occupies a central role (similar to say the MQ problem). Hence, there exist a number of nontrivial attacks on this problem, as well as varied problems where it plays a central role in their cryptanalysis, and finally also a number of cryptographic schemes where it is used as the underlying hardness assumption. 
It is instructive to look at the history of MinRank following these separate themes. We present a summary below.

\paragraph{Usage in cryptanalysis:} The MinRank problem comes up naturally in the cryptanalysis of certain schemes in multivariate cryptography (for example, by inherently capturing certain key recovery attacks for such cryptosystems). This has happened with the HFE cryptosystem and variants \cite{KS99,Courtois01b,FaugereJ03,CourtoisDF03,BettaleFP11,TaoPD21,BaenaBCPSV22} and the TTM cryptosystem \cite{goubinCryptanalysisTTMCryptosystem2000}. \cite{CSV97} developed what would later turn into the minors attack as a component in their attack on birational permutation signature schemes. \cite{MoodyPS14} and \cite{MoodyPS17} use a MinRank based approach to attack the ABC matrix encryption scheme and simple cubic encryption scheme respectively. \cite{BriaudTV21} exploit MinRank attacks to completely break the multivariate encryption scheme of \cite{RavivLT21} relying on Sidon spaces (which are special base-field subspaces of an extension field).  

The MinRank problem was also shown to be linked closely to the Rank Decoding problem (in \cite{Courtois06,faugereCryptanalysisMinRank2008}), which is essentially the syndrome decoding problem for the so-called {\em rank-metric codes} or rank codes. These are error correcting codes defined over extensions of some finite field, which then allows for the codewords to be represented as matrices over the base field. One can then define the usual notions of error correction viewing the rank difference between two matrices as the relevant metric. We discuss cryptography based on such codes briefly at the beginning of \cref{sec:intro}

In particular, for an excellent overview of such codes, see \cite{bartzRankMetricCodesTheir2022}.  

As described previously, these codes (as well as the underlying hardness of decoding) have been used to design McEliece-style cryptosystems (discussed also in \Cref{sec:dual}, and also very extensively in \cite{bartzRankMetricCodesTheir2022}), culminating in the schemes RQC \cite{ABDGZ16,bidoux2022rqc} and ROLLO \cite{rollo2020nist} which were notable submissions for the NIST PQC competition for selecting and standardizing post-quantum cryptosystems (\cite{nist2016pqcstandardization}). Attacks on MinRank lead to ones on Rank Decoding, and thus play a major role in the cryptanalysis of such systems. In particular, the security of suggested parameter sets for both RQC and ROLLO were  broken comprehensively by \cite{bardetAlgebraicAttackRank2020,bardetImprovementsAlgebraicAttacks2020a}, and the schemes were subsequently withdrawn from the NIST competition (we describe these attacks in more detail below).

More recently, there have been successful attacks on the GeMSS  and RAINBOW signature schemes by \cite{BaenaBCPSV22} and \cite{Beullens22}, respectively, that rely on MinRank attacks. The latter scheme was also a NIST PQC candidate that reached the final stages of evaluation till it was broken comprehensively by \cite{Beullens22}. 

\paragraph{Attacks on MinRank:} Notable attacks include the one by Kipnis and Shamir \cite{KS99}, the kernel attack \cite{goubinCryptanalysisTTMCryptosystem2000}, and the minors attack \cite{Courtois01b}. \cite{faugereCryptanalysisMinRank2008} showed a new formalism based on the Kipnis-Shamir framework, gave direct complexity estimates of their attack, and showed that it is efficient in certain parameter settings. They also showed that the KS framework is as exhaustive as the Minors attack. \cite{JiangDH08} showed that the Kipnis-Shamir and minors attacks are unlikely to be efficient in the context of HFE cryptanalysis. 

\cite{faugereCryptanalysisMinRank2008} gave an improved analysis of the minors attack for the case of square matrices, linking it to the theory of determinantal ideals. They also gave complexity estimates that suggest that this attack does slightly better than the Kipnis-Shamir attack, and gave a range of parameters where the minors attack is efficient (in particular, this includes setting the rank being either a small constant or a small constant away from the matrix dimension $n$). \cite{faugereComplexityGeneralizedMinRank2013} carried this analysis to the more general setting of MinRank with rectangular matrices. \cite{VBCPS19} showed an improved analysis of the Kipnis-Shamir attack combined with the XL algorithm for what they called the `superdetermined' setting, which remarkably improved the previous known complexity of the attack. \cite{NakamuraWI23} extended their approach to work for a slightly wider parameter setting. \cite{GND23} provided an improved, more efficient \groebner basis algorithm that is optimized specifically for the minors-based modeling of MinRank.  


Building on the works of \cite{VBCPS19} and \cite{bardetAlgebraicAttackRank2020}, the work of \cite{bardetImprovementsAlgebraicAttacks2020a} developed the Support-Minors attack, which showed a significant improvement over the previous attacks for a wide range of parameters. They also gave a related attack for rank decoding. A slew of subsequent works provide improved rigorous analysis (\cite{BG25,BardetB22,BBBGT22,CGG25}), widen applicability of the attack (\cite{BBBGT22,BaenaBCPSV22}), clarify links to other existing attacks (\cite{BardetB22,GuoD22}), and find additional applications to cryptanalysis (\cite{BaenaBCPSV22,Beullens22}). In particular, the attack led to showing weaknesses in the suggested parameter sets for RQC and ROLLO, and a variant of it in \cite{Beullens22} was used as the attack of choice to break RAINBOW.  \cite{AdjBBERSVZ24} give a quantum algorithm based on the Kipnis Shamir modeling and Grover search that gives a quadratic speedup over the standard Kernel attack. 

\paragraph{Cryptographic usage:} There have been a number of cryptographic  constructions over the years relying on hardness of the MinRank assumption. \cite{Cou01} constructed a 3-round identification scheme with zero knowledge (or a sigma protocol in modern parlance), and combined it with the Fiat-Shamir heuristic to obtain a signature scheme. This remained the only such construction for a number of years, until recent work. 

\cite{MD22} gave an improved sigma protocol with reduced soundness error, and an efficient signature scheme based on the sigma protocol with helpers paradigm. \cite{Feneuil24,AdjRV23} also give improved signature schemes using the MPC in the head paradigm to get problem specific arguments of knowledge which have been increasingly used in recent years in the design of efficient signatures. 

\cite{ABCFGNR23} and \cite{AdjBBERSVZ24} constructed even more efficient signature schemes, building on the work of \cite{Feneuil24} and \cite{AdjRV23} respectively. These were later combined into the MiRath scheme \cite{adj2024mirath} that was submitted to the NIST PQC standardisation contest for post-quantum signatures. All of the schemes mentioned above can be also modified to obtain ring signatures, using standard techniques (as mentioned in \cite{Cou01,MD22,adj2024mirath}). 

All of the above works rely on the hardness on average of the Search MinRank problem. Such hardness is justified using an approach similar to ours, wherein the most prominent attacks are examined and their complexities are estimated, focusing on how these attacks behave on random instances of MinRank. Interestingly, the work of \cite{SantosoINY22} construct a sigma protocol improving on \cite{Cou01}, based on the decisional MinRank problem. While they formulate this as a separate assumption, they do not assess its hardness beyond noting that the problem is NP-hard in the worst case setting (which was observed in \cite{BFS99}).    
